\documentclass[twocolumn, prb,showpacs]{revtex4-2}
\usepackage{graphicx}
\usepackage{amssymb}
\usepackage{amsmath}
\usepackage{graphicx}
\usepackage{bm}
\usepackage{hyperref}
\usepackage{subcaption}
\usepackage{dcolumn}

\begin{document}
\title{Electronic Structure of Multilayer Graphene with Arbitrary Stackings}
\author{Fred Sun}
\email{fredsunqiushi@gmail.com}
\affiliation{BASIS Independent Silicon Valley, 1290 Parkmoor Ave, San Jose, CA 95126}
\author{Jia-An Yan}
\affiliation{Department of Physics, Astronomy, and Geosciences, Towson University
8000 York Road, Towson, MD 21252, USA}
\date{\today}
\email{jiaanyan@gmail.com}

\begin{abstract}
Stacking geometry in multilayer graphene (MLG) provides an interesting degree of freedom to engineer its electronic structure near the Fermi level, wherein the linear bands in single layer graphene could retain or evolve into parabolic or flat bands. Using a tight-binding model, we carried out a detailed analytical analysis of the electronic band structures for arbitrarily stacked MLGs. We show that their low energy band dispersions near the Fermi level may be deduced from its substacks in isolation. The analytical solutions of the momenta with zero eigenvalue for an AA stacking allows us to generalize the results of the zero energy momenta for arbitrarily stacked MLGs. Moreover, we find that an interplay of parallel and rhombohedral stackings allows for flat band engineering and enhancement in arbitrarily stacked MLGs. The existence of flat bands in MLGs might offer another interesting platform for exploring the superconductivity in graphene systems beyond the twisted bilayer graphene.
\end{abstract}

\maketitle

\section{Introduction}

The electronic properties of multilayer graphene are fundamentally determined by its stacking geometry, which controls the band structure near the Fermi level. While single layer graphene features massless Dirac fermions with a linear dispersion \cite{experimental}, multilayer graphene systems exhibit dramatically different behaviors depending on their stacking order and layer number. Early theoretical and experimental studies established that Bernal (AB), rhombohedral (ABC), and parallel (AA) stackings each produce distinct electronic structures \cite{Latil2006FLG,Castro_Neto,Yan,Macdonald}. Moreover, the nature of charge carriers in few-layer graphene is highly sensitive to both the number of layers and the stacking geometry, making the possibility of analyzing a general arbitrary multilayer stack challenging. Rhombohedral stacking is particularly notable for generating flat bands at low energies, characterized by vanishing Fermi velocity and enhanced density of states. These flat bands play a central role in promoting strong electron-electron correlations and have become a key ingredient for realizing exotic quantum phases including superconductivity \cite{superconductor}.

The discovery of superconductivity in magic-angle twisted bilayer graphene has generated intense interest in flat band engineering as a route to unconventional superconductivity \cite{bandtwist,mactwist}. When graphene layers are twisted at specific magic angles (the first  being $\theta \approx 1.1$\textdegree), moir\'e superlattices form with isolated flat bands that support Mott insulating states and superconducting domes reminiscent of high-$T_c$ cuprates. This paradigm has been extended to multilayer systems; for example, magic-angle twisted multilayer graphene ($N$ = 3, 4, 5 layers) forms a robust family of moir\'e superconductors, revealing that flat bands are a common feature but with important variations between even and odd layer numbers \cite{Park2021MAG}. Moreover, gate-tunable  superconductivity in ABC-trilayer graphene/hexagonal boron nitride systems have been observed, showing Mott insulating states and superconducting domes that can be continuously tuned via displacement fields \cite{Chen2019TLG}. More remarkably, superconductivity has been observed in rhombohedral trilayer graphene without moir\'e effects, demonstrating that crystalline stacking order alone can generate the necessary conditions for superconductivity \cite{Zhou2021RTG}. Most recently, signatures of chiral superconductivity in rhombohedral tetra- and penta-layer graphene with transition temperatures up to 300 mK, featuring time-reversal symmetry breaking and magnetic hysteresis hallmarks of unconventional pairing has been observed \cite{Han2024Chiral}. Hence, multilayer graphene has emerged as a versatile platform for exploring correlated electron physics and unconventional superconductivity. To access the full spectrum of electronic properties multilayer graphene offers, however, requires an analysis of completely arbitrary stacking. In doing so, one could predict and engineer a generalized stack to fit their electronic purposes. Understanding how different stacking configurations produce flat bands and influence the electronic properties at the Fermi level is therefore crucial for identifying and designing new foundations for correlated phenomena and superconductivity in graphene systems.

In this work, we carry out an analytical and computational study of the electronic band structures of arbitrarily stacked multilayer graphene (MLG) using a tight-binding model. Stacking sequence in MLG offers a degree of freedom to tune its electronic band dispersions. Within nearest-neighbor interactions (hopping), our Hamiltonian takes on a relatively simple form and the low energy band dispersions near the $K$ and $K'$ points are calculated. The solutions to pure parallel (AA) and Bernal (AB) stackings, along with the system of (transcendental) equations for ABC stacking, will be derived first. We will then investigate MLG with stacking fault by embedding a certain type of pure stack (e.g., AA) into another type of pure stack (e.g., AB). For MLGs with pure AA substacks, the linear bands persist with the crossing points shifted to other k-points other than $K$ and $K'$. We find an analytical solution to predict the locations of these k-points. Furthermore, we will show the robustness of the flat bands in arbitrarily stacked MLGs, which are related to the embedded ABC substacks. The obtained electronic band dispersions are further confirmed by first principle calculations. Our study shows that the electronic properties of MLGs with an arbitrary stacking sequence may be deduced from its substacks considered in isolation, enabling an interesting design of the desired electronic structures in MLGs.

The paper is organized as follows: In Section \ref{sect:method}, we briefly introduce the tight binding model used in this work. In Section \ref{sect:result}, we present detailed results for single layer graphene and multilayer graphene. A conclusion is given in Section \ref{sect:conclusion}.

\section{Method} \label{sect:method}

We use a tight binding method with nearest neighbor hopping, treating each atomic site as a basis state and constructing our Hamiltonian so that if $i$ and $j$ are neighboring sites then $\langle i|H|j\rangle = t$ and all non-neighboring hopping integrals are zero \cite{Kittel}. This method has been well received for producing accurate results for many physical systems \cite{SSH} \cite{rutgers}. For graphene, we only consider the interactions between the $2p_z$ orbitals, since they give rise to the strongest hopping. The simulation parameters normalize $\hbar = 1$ and lattice constant $a = 1$, with $t = 3$ (approximately its value in $eV$). For MLGs, we only consider the interlayer hopping when two carbon atoms are on top of each other. We set this interlayer hopping term $t_\perp = \frac{t}{10} = 0.3$. 

First-principle calculations based on density-functional theory (DFT) are also performed using Quantum Espresso (QE) code \cite{pwscf} with local density approximation (LDA). Norm-conserving psuedopotentials for carbon has been adopted. A 36$\times$36$\times$1 Monkhorst-Pack uniform $k$-grid and cutoff energy of 90 Ryd have been employed.
\section{Results and Discussions} \label{sect:result}

We will first review the basic properties of single layer graphene and certain types of graphene stacking. We then investigate the MLGs with an arbitrary stacking sequence.

\subsection{Single Layer Graphene}

\begin{figure}[tbp]
    \centering
    \setkeys{Gin}{width=\linewidth}
    \begin{subfigure}{.45\textwidth}
        \includegraphics{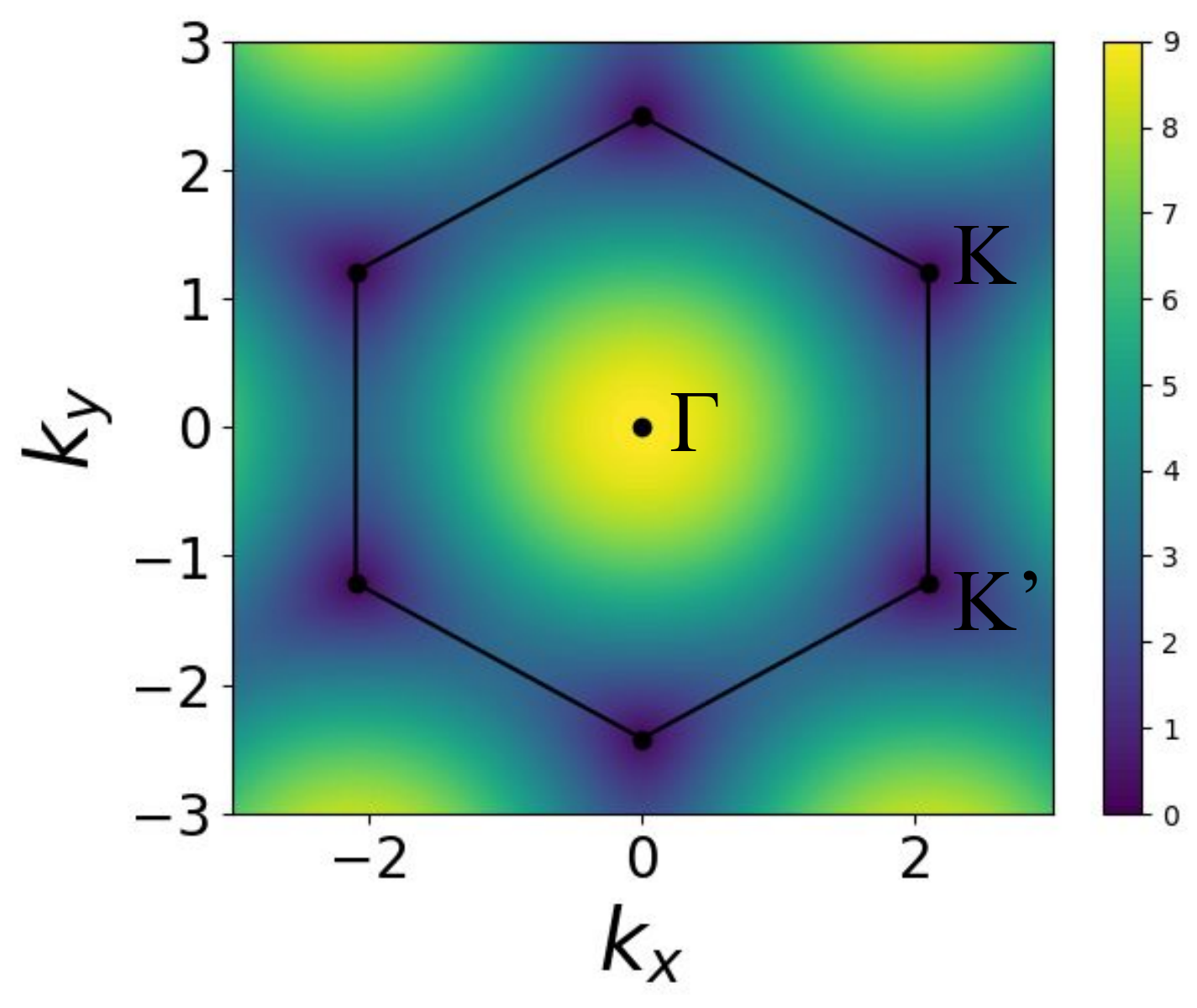}
        \caption{}
        \label{fig:1a}
    \end{subfigure}
    \hfil
    \begin{subfigure}{.45\textwidth}
        \includegraphics{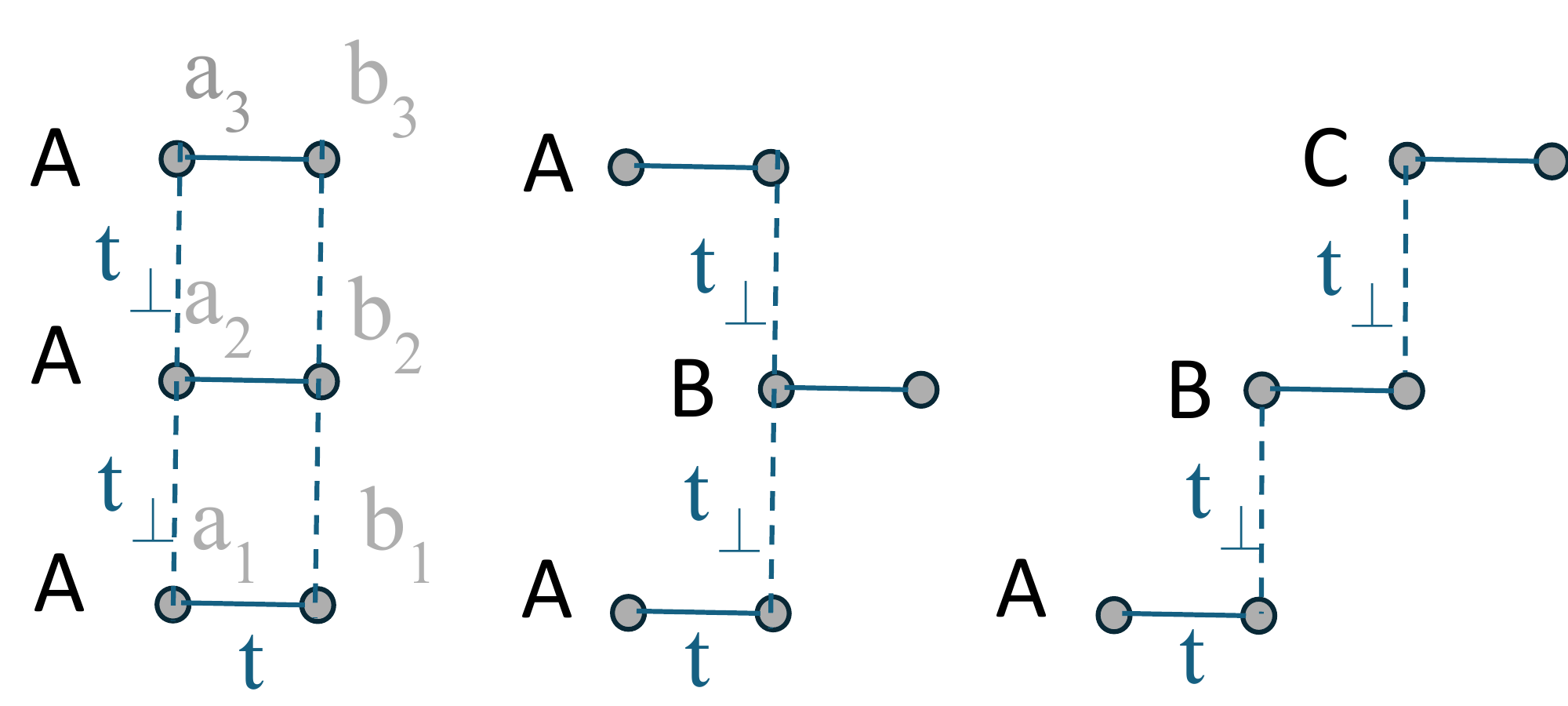}
        \caption{}
        \label{fig:1b}
    \end{subfigure}
   \captionsetup{justification=raggedright, singlelinecheck=false}
   \flushleft \caption{(Color online)(a) Contour map of the positive branch of the band dispersion for monolayer graphene in the whole hexagonal Brillouin zone. The corners of the hexagon are the points of zero energy. Examples of the $K$ and $K'$ points have been shown. (b) Schematic of AAA, ABA, and ABC graphene stacking geometry in the Bloch basis. The $a$ and $b$ bases (gray), hopping integrals (cyan), and stacking letters (black) have been shown, which naturally constructs the Hamiltonian.}\label{fig1}
\end{figure}

As a hexagonal lattice, monolayer graphene has a primitive unit cell of two carbon atoms denoted $a$ and $b$. Each carbon atom has three nearest-neighbor interactions; we label nearest-neighbor atoms to have electron annihilation operators $a_{\vec{R}}$ and $b_{\vec{R}+\vec{r_i}}$, where $\vec{r_i}$ ($i$ =1-3) denotes vectors to its three nearest neighbors, so that the Hamiltonian may be constructed as \cite{crystal_structure}
\begin{equation}
    H = -t\sum_{\vec{R},i}a^\dagger_{\vec{R}}b_{\vec{R}+\vec{r_i}}+h.c.
\end{equation}
Converting to the Bloch bases with annihilation operators for a given momentum $a_{\vec{k}}$ and $b_{\vec{k}}$ defined by
\begin{equation}
  a_{\vec{R}} = \frac{1}{\sqrt{N}}\sum_{\vec{k}}a_{\vec{k}}e^{i\vec{k}\cdot\vec{R}}, b_{\vec{R}} = \frac{1}{\sqrt{N}}\sum_{\vec{k}}b_{\vec{k}}e^{i\vec{k}\cdot\vec{R}},
\end{equation}
gives us \cite{tutorial}
\begin{equation}
  H = -t\sum_i\sum_{\vec{k}}a^\dagger_{\vec{k}}b_{\vec{k}}e^{i\vec{k}\cdot\vec{r_i}} + h.c.,
\end{equation}
which in matrix form gives
\begin{equation}
    H(\vec{k}) = \begin{pmatrix} 0 & -t\sum_ie^{i\vec{k}\cdot\vec{r_i}}\\ -t\sum_ie^{-i\vec{k}\cdot\vec{r_i}} & 0 \end{pmatrix}.
\end{equation}
Since we are working for a given $\vec{k}$, we will omit explicitly giving the $\vec{k}$ dependence in our bases (there is, of course, $\vec{k}$ dependence on our Hamiltonian components however) and refer to the Bloch bases as simply $a$ and $b$. The dispersion relation for $\vec{k} = \begin{pmatrix}k_x \\ k_y\end{pmatrix}$ can be shown to be \cite{Semenoff,Wallace}
\begin{equation}
    \resizebox{0.4\textwidth}{!}{$\epsilon_{\pm, \vec{k}} = \pm t\sqrt{3 + 2\cos\left(\sqrt{3}k_y a\right) + 4\cos\left(\frac{\sqrt{3}k_y a}{2}\right)\cos\left(\frac{3k_x a}{2}\right)}$},
\end{equation}
which is plotted in Fig.~\ref{fig:1a}. Note that how the low energy regions (purple) form a hexagon. We have sketched the points of zero energy, which are also the corners of the first Brillouin Zone. We may choose $K$ and $K'$ as shown in the figure so that all the other zero energy points are connected to these two points by reciprocal lattice vectors.

For monolayer graphene, the two bands cross at the Fermi level, suggesting that low energy analysis should be done near the $K$ or $K'$ points. We will choose $K$. Specifically, we shift the origin to be at $\vec{K}$ and measure the momentum relative to $\vec{K}$, which we will call $\vec{p} = \vec{k} - \vec{K}$. Performing a linear expansion $h(\vec{K}+\vec{p})$ we will, defining $\vec{p} = \begin{pmatrix}p_x \\ p_y\end{pmatrix}$, the Fermi velocity $v = \frac{3ta}{2}$, and $\pi = p_x + ip_y$, arrive at, upon performing a global gauge transformation, \cite{tutorial}
\begin{equation}
    H(\vec{p}) = \begin{pmatrix} 0 & v\pi^\dagger\\ v\pi& 0 \end{pmatrix},
\end{equation}
where we recognize the 2D massless Dirac Hamiltonian \cite{Castro_Neto}:
\begin{equation}
    H = v\boldsymbol{\sigma}\cdot\vec{p}.
\end{equation}
The dispersion relation, which is
\begin{equation}
    \epsilon_{\pm,\vec{p}} = \pm v|\vec{p}|,\label{eq:massless_dirac_fermions}
\end{equation}
is notably linear. This is consistent with the band structure of the exact dispersion relation near the $K$ and $K'$ points.
\subsection{Multilayer Graphene}
\label{sec:stacking}
Graphene stacking is characterized by how we orient each sheet relative to the one below it, which working with $a$ and $b$ is understood as where we stack each successive $a_{i+1}$ and $b_{i+1}$ sites onto the $a_i$ and $b_i$ sites. We have three options. We may orient the $a_{i+1}$ site onto the $b_i$ site; or orient each $a_{i+1}$ and $b_{i+1}$ onto $a_i$ and $b_i$ respectively; or orient $b_{i+1}$ onto $a_i$. These three choices give rise to three stacking orientations A, B, and C, as shown in Fig.~\ref{fig:1b}. In general, we can choose the first stack in our sequence to be of type A, which is done without loss of generality. Then, among the above stacking options, AB, BC, and CA all refer to the first stacking option; AA, BB, and CC all refer to the second stacking option; AC, CB, and BA all refer to the third stacking option. Three different stackings for trilayer graphene have been sketched in Fig.~\ref{fig:1b}. We will first consider a general, arbitrary stack and then apply it to the well known cases. For a given momentum, we omit the $\vec{k}$ subscripts in our creation and annihilation operators (like we did for the Bloch bases), assuming it to be implied. We thus have
\begin{widetext}
\begin{equation}
      H = (v\pi^\dagger\sum_{i=1}^N b^\dagger_{i} a_{i} + t_\perp\sum_{j=1}^{N-1} (\alpha_ja^\dagger_{j+1}b_{j} + \beta_j(a^\dagger_{j+1}a_{j} + b^\dagger_{j+1}b_{j}) + \gamma_jb^\dagger_{j+1}a_{j})) + h.c..
\end{equation}
In matrix form, we have
\begin{equation}
  H(\vec{p}) = t_\perp\begin{pmatrix}
  0 & f\pi^{\dagger} & \beta_1 & \gamma_1 & 0 & 0 & \cdots & 0 & 0 \\
  f\pi & 0 & \alpha_1 & \beta_1 & 0 & 0 & \cdots & 0 & 0 \\
  \beta_1 & \alpha_1 & 0 & f\pi^{\dagger} & \beta_2 & \gamma_2 & \cdots & 0 & 0 \\ \gamma_1 & \beta_1 & f\pi & 0 & \alpha_2 & \beta_2 & \cdots & 0 & 0 \\
  0 & 0 & \beta_2 & \alpha_2 & 0 & f\pi^{\dagger} & \cdots & 0 & 0 \\
  0 & 0 & \gamma_2 & \beta_2 & f\pi & 0 & \cdots & 0 & 0 \\
  \vdots & \vdots & \vdots & \vdots & \vdots & \vdots & \ddots & \vdots & \vdots \\
  0 & 0 & 0 & 0 & 0 & 0 & \cdots & 0 & f\pi^{\dagger} \\
  0 & 0 & 0 & 0 & 0 & 0 & \cdots & f\pi & 0
\end{pmatrix},
\end{equation}
\end{widetext}
where $f = \frac{v}{t_\perp}$ and we have defined $(\alpha_i,\beta_i,\gamma_i)$ to follow the relations:
\begin{equation}
  \alpha_j\beta_j = 0
\end{equation}
\begin{equation}
  \beta_j\gamma_j = 0
\end{equation}
\begin{equation}
  \gamma_j\alpha_j = 0
\end{equation}
\begin{equation}
  \alpha_j + \beta_j + \gamma_j = 1.
\end{equation}
In other words $(\alpha_j, \beta_j, \gamma_j) \in \{(1,0,0), (0,1,0), (0,0,1)\}$, wherein AB, BC, CA stacking refers to the $(1,0,0)$ solution; AA, BB, CC stacking refers to the $(0,1,0)$ solution; and BA, CB, AC stacking refers to the $(0,0,1)$ solution. There are three well studied stacking structures. The first we consider is AA stacking, what we will call parallel stacking, where we have a stacking sequence AAAA... given by the Hamiltonian
\begin{equation}
  H_{AA} = \begin{pmatrix}
  0 & v\pi^{\dagger} & t_{\perp} & 0 & 0 & 0 & \cdots & 0 & 0 \\
  v\pi & 0 & 0 & t_{\perp} & 0 & 0 & \cdots & 0 & 0 \\
  t_{\perp} & 0 & 0 & v\pi^{\dagger} & t_{\perp} & 0 & \cdots & 0 & 0 \\
  0 & t_{\perp} & v\pi & 0 & 0 & t_{\perp} & \cdots & 0 & 0 \\
  0 & 0 & t_{\perp} & 0 & 0 & v\pi^{\dagger} & \cdots & 0 & 0 \\
  0 & 0 & 0 & t_{\perp} & v\pi & 0 & \cdots & 0 & 0 \\
  \vdots & \vdots & \vdots & \vdots & \vdots & \vdots & \ddots & \vdots & \vdots \\
  0 & 0 & 0 & 0 & 0 & 0 & \cdots & 0 & v\pi^{\dagger} \\
  0 & 0 & 0 & 0 & 0 & 0 & \cdots & v\pi & 0
\end{pmatrix}.
\end{equation}

\begin{figure}[tbp]
    \centering
\setkeys{Gin}{width=\linewidth}
\begin{subfigure}{.45\textwidth}
  \includegraphics{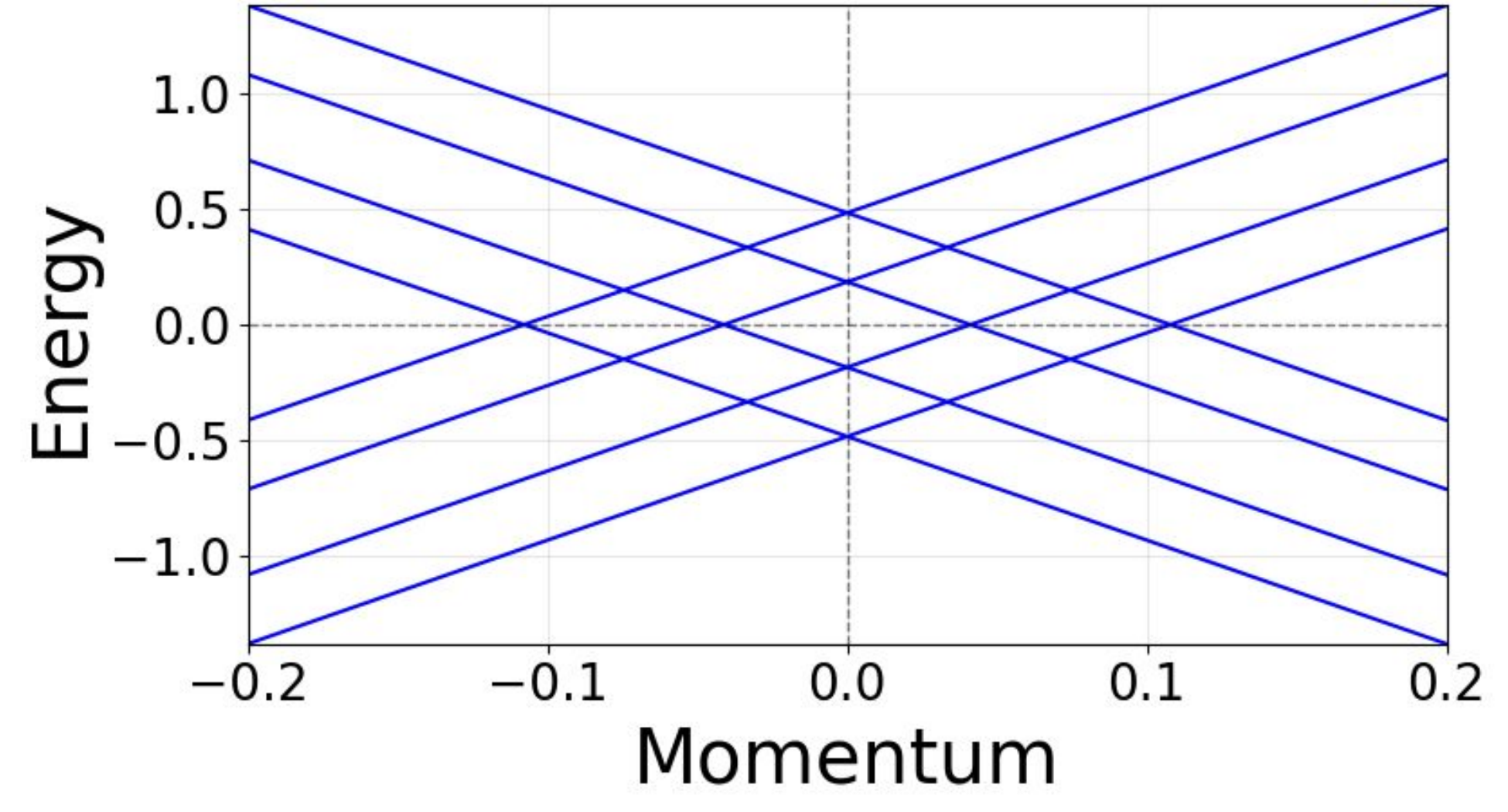}
  \caption{}
    \label{fig:2a}
\end{subfigure}
\hfil
\begin{subfigure}{.45\textwidth}
  \includegraphics{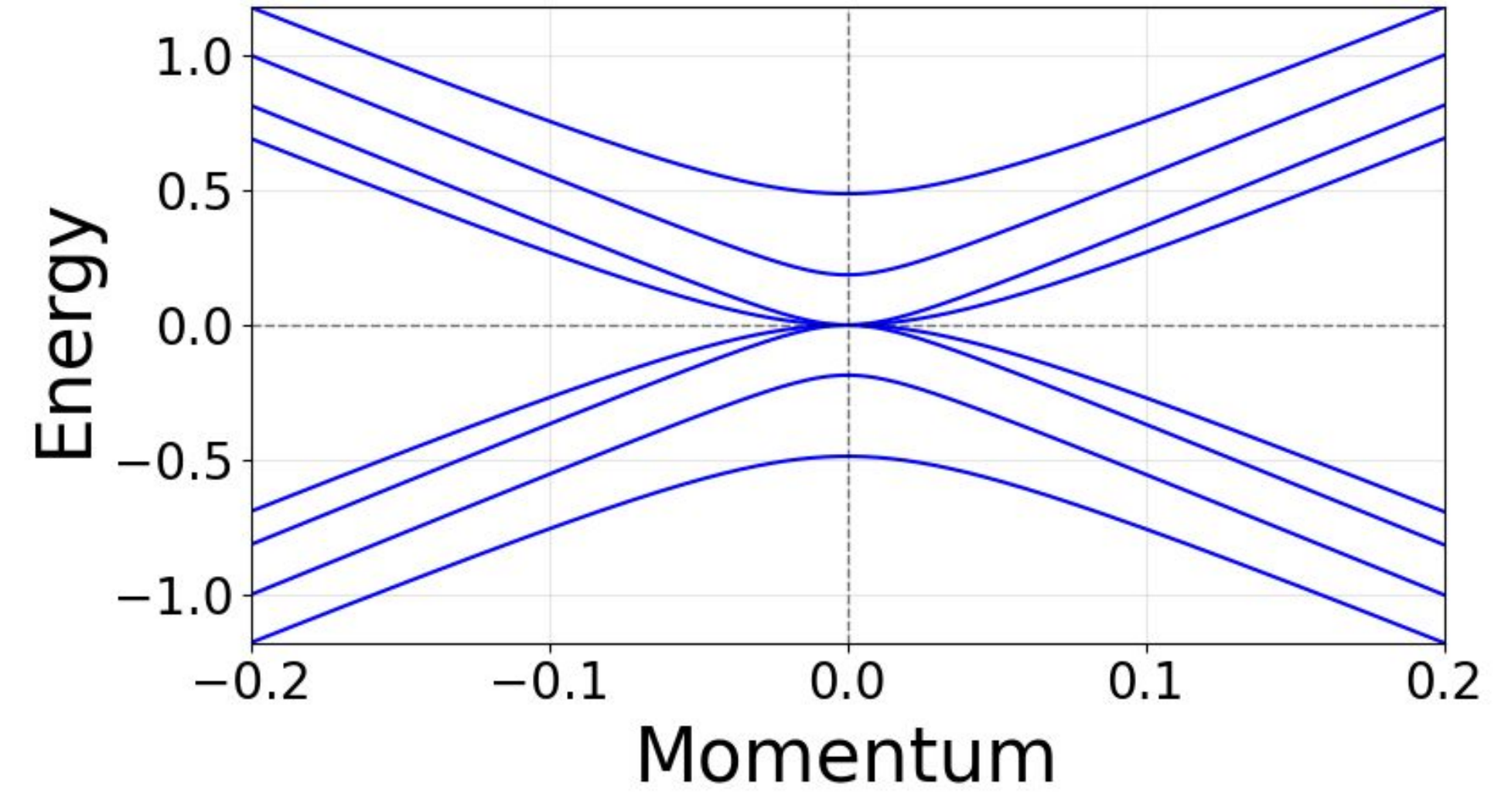}
  \caption{}
\label{fig:2b}
\end{subfigure}

\begin{subfigure}{.45\textwidth}
  \includegraphics{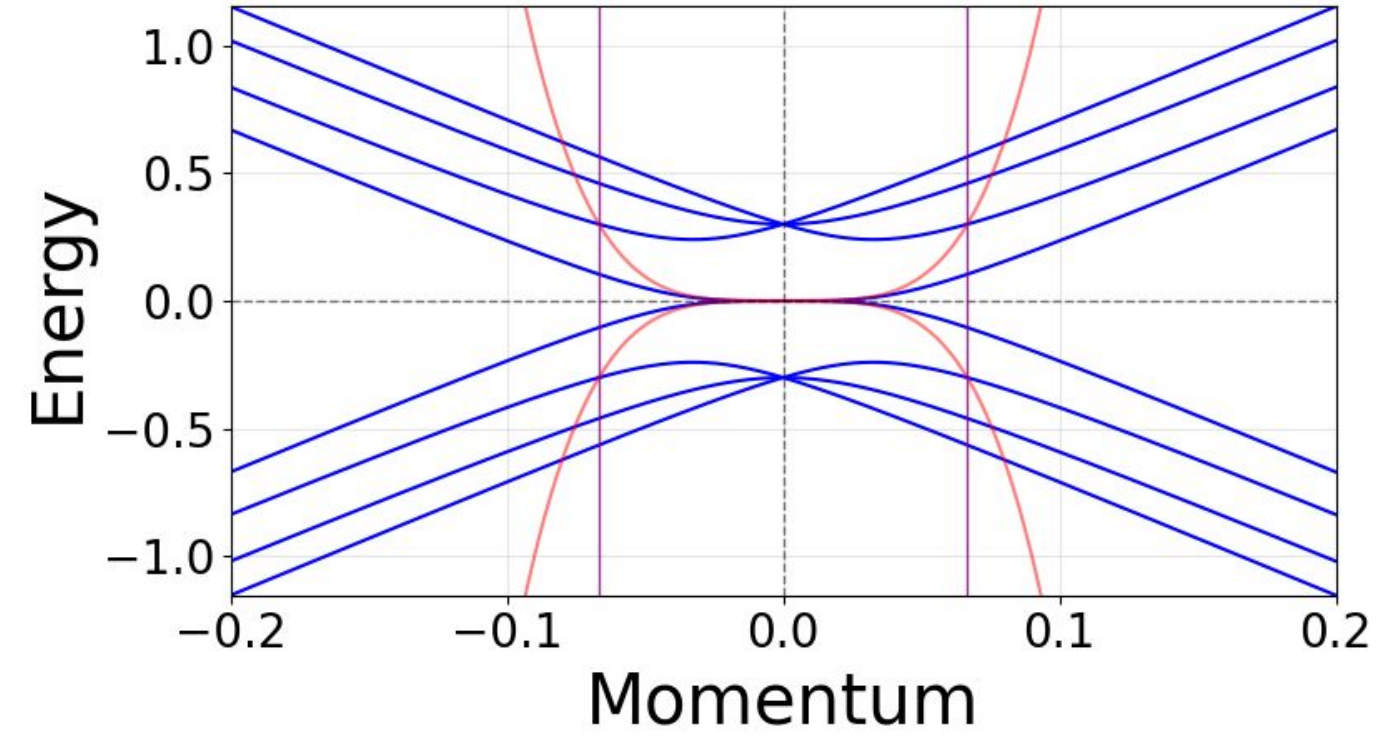}
  \caption{}
  \label{fig:2c}
\end{subfigure}
\captionsetup{justification=raggedright, singlelinecheck=false}
\caption{Calculated electronic band dispersions for (a) tetralayer AA stacked graphene, (b) tetralayer AB stacked graphene, and (c) tetralayer ABC stacked graphene. The approximate low-energy band analytical solution (red) are also shown for comparison in (c). Note the accuracy at small momentum. The maximum flat band range, when $N$ is large, is bounded by the momentum values plotted via purple vertical lines.}
\label{tetralayer-dispersion-relations}
\end{figure}

Upon employing recurrence methods we reach the dispersion relation \cite{Macdonald}
\begin{equation}
  \epsilon^{\pm,AA}_{r,\vec{p}} = \pm v|\vec{p}| + 2t_\perp \cos(\frac{r\pi}{N+1}) \label{eq:AA_disp},
\end{equation}
where $r \in \{1,2,3...N\}$ and $N$ is the length (number of layers) of our stack. In general, an $N$-layer stack will have a Hamiltonian with dimensions of $2N\times 2N$. We notice that this is simply a sum of the eigenenergies of the Dirac Hamiltonian and that of the open chain tight-binding model \cite{Kittel}. It is intuitive due to the parallel nature of the stack. Moreover, it will be useful later for comparison to find specifically the momentum values $|\vec{p}|$ for which $\epsilon^{\pm,AA}_{r,\vec{p}} = 0$. They are given by
\begin{equation}
    |\vec{p}| = |\frac{2t_\perp}{v}\cos(\frac{r\pi}{N+1})|, \label{eq:zero_AA_eigen}
\end{equation}
where the absolute value on the cosine function is justifiable since our dispersion relation concerns only with $\pm|\vec{p}|$. The dispersion relation for tetralayer (4-layer) AA stacking is plotted in Fig.~\ref{fig:2a}. Notice the linear bands that resemble the shifted dispersion relation for monolayer massless Dirac fermions. These shifts generate multiple band crossings at momentum values other than the $\vec{K}$ and $\vec{K'}$ points. These crossings are characteristic of AA stacking, a property that will be insightful in analyzing arbitrary stacking.

For AB stacking, which is referred to as Bernal stacking, we have a stacking sequence ABABAB... which is given by the Hamiltonian
\begin{equation}
  H_{AB} = \begin{pmatrix}
  0 & v\pi^{\dagger} & 0 & 0 & 0 & 0 & \cdots & 0 & 0 \\
  v\pi & 0 & t_{\perp} & 0 & 0 & 0 & \cdots & 0 & 0 \\
  0 & t_{\perp} & 0 & v\pi^{\dagger} & 0 & t_{\perp} & \cdots & 0 & 0 \\
  0 & 0 & v\pi & 0 & 0 & 0 & \cdots & 0 & 0 \\
  0 & 0 & 0 & 0 & 0 & v\pi^{\dagger} & \cdots & 0 & 0 \\
  0 & 0 & t_{\perp} & 0 & v\pi & 0 & \cdots & 0 & 0 \\
  \vdots & \vdots & \vdots & \vdots & \vdots & \vdots & \ddots & \vdots & \vdots \\
  0 & 0 & 0 & 0 & 0 & 0 & \cdots & 0 & v\pi^{\dagger} \\
  0 & 0 & 0 & 0 & 0 & 0 & \cdots & v\pi & 0
  \end{pmatrix}.
\end{equation}

Upon employing similar recurrence methods for this matrix, we can reach the dispersion relation \cite{Macdonald}
\begin{equation}
  \epsilon^{\pm,AB}_{r,\vec{p}} = t_\perp \cos(\frac{r\pi}{N+1}) \pm \sqrt{v^2|\vec{p}|^2 + t_\perp^2 \cos^2(\frac{r\pi}{N+1})}. \label{eq:AB_disp}
\end{equation}
This equation for tetralayer AB stacking is plotted in Fig.~\ref{fig:2b}. The structure of the equation follows the massive Dirac fermion dispersion relation, with parabolic bands at low energies.

Finally, the ABC stacking, denoted by a stacking sequence ABCABCABC..., is given by the Hamiltonian
\begin{equation}
  H_{ABC} = \begin{pmatrix}
  0 & v\pi^{\dagger} & 0 & 0 & 0 & 0 & \cdots & 0 & 0 \\
  v\pi & 0 & t_{\perp} & 0 & 0 & 0 & \cdots & 0 & 0 \\
  0 & t_{\perp} & 0 & v\pi^{\dagger} & 0 & 0 & \cdots & 0 & 0 \\
  0 & 0 & v\pi & 0 & t_{\perp} & 0 & \cdots & 0 & 0 \\
  0 & 0 & 0 & t_{\perp} & 0 & v\pi^{\dagger} & \cdots & 0 & 0 \\
  0 & 0 & 0 & 0 & v\pi & 0 & \cdots & 0 & 0 \\
  \vdots & \vdots & \vdots & \vdots & \vdots & \vdots & \ddots & \vdots & \vdots \\
  0 & 0 & 0 & 0 & 0 & 0 & \cdots & 0 & v\pi^{\dagger} \\
  0 & 0 & 0 & 0 & 0 & 0 & \cdots & v\pi & 0
  \end{pmatrix}.
\end{equation}

The analytical solution can be represented as a system of two transcendental equations \cite{Yan}. One of the characteristics of the Hamiltonian, however, is the localized eigenstate that produces a low energy flat band. We may apply a Schrieffer-Wolf transformation to the Hamiltonian to obtain the low energy dispersion relation \cite{Schrieffer-Wolff}\cite{Macdonald}
\begin{equation}
    \epsilon^{\pm,ABC}_{\vec{p}} = \pm t_\perp(\frac{v|\vec{p}|}{t_\perp})^N.
\end{equation}

Since the Schrieffer-Wolf transformation is derived from assuming that the energy eigenvalue is much smaller than the elements of the Hamiltonian matrix, define $\Delta|\vec{p}|$ such that
\begin{equation}
    t_\perp(\frac{v\Delta|\vec{p}|}{t_\perp})^N = t_\perp.
\end{equation}
The momenta around and greater than $\Delta|\vec{p}|$ is when the Schrieffer-Wolf transformation becomes inaccurate. So, the radius in which this approximation is accurate is given by
\begin{equation}
    \Delta|\vec{p}| = \frac{t_\perp}{v}. \label{eq:ABC_band_radius}
\end{equation}
Moreover, we could also have derived this as $\lim\limits_{N\to\infty}(\frac{v\Delta|\vec{p}|}{t_\perp})^N = 1$, which shows that Eq. \eqref{eq:ABC_band_radius} also approximates  the radius of the flat band in the large layer number $N$ limit. In this limit, at $|\vec{p}| < \Delta|\vec{p}|$, our eigenenergy is almost zero, and at $|\vec{p}>\Delta|\vec{p}|$, our eigenenergy becomes very large. Note that the localization of this flat band is specifically observed for trilayer ABC graphene and above; we understand that bilayer rhombohedral graphene and bilayer Bernal graphene are the same. Fig. ~\ref{fig:2c} shows the dispersion relation for tetralayer ABC graphene. Clearly, the flat bands near the Fermi level are present. We have also plotted our approximated flat band dispersion relation in red, which achieves accuracy at low momentum (and therefore low energy) values. As our layer number $N$ becomes large, our approximation also gives a maximum momentum radius of our flat band according to Eq. \ref{eq:ABC_band_radius}, which we have plotted in purple.

\begin{figure}[tbp]
    \centering
    \includegraphics[scale = 0.3]{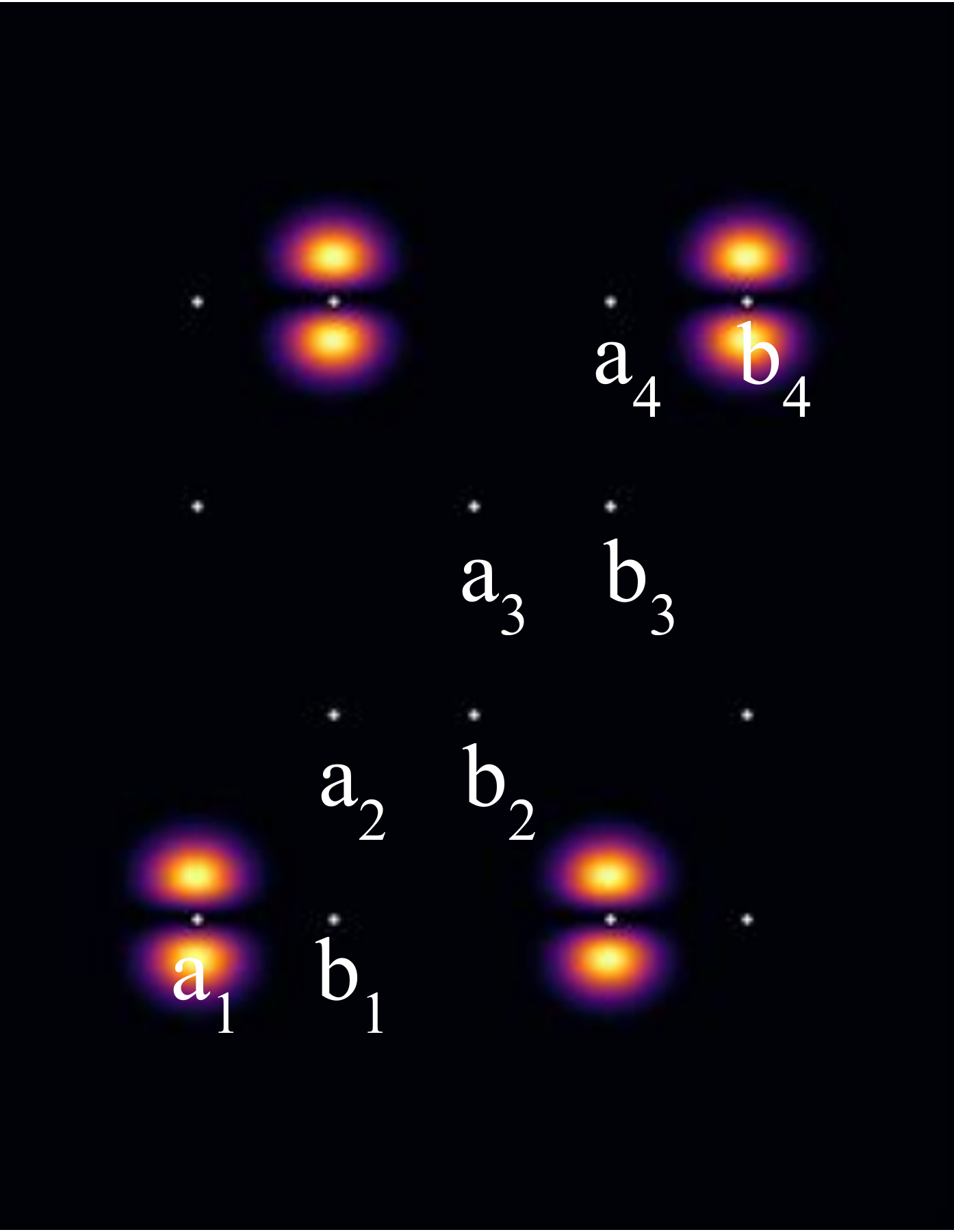}
    \captionsetup{justification=raggedright, singlelinecheck=false}
    \caption{The low energy charge density for tetralayer ABC stacked graphene with sufficiently low momentum in the flat band. The stack is copied over horizontally to show the periodic nature of the crystal lattice. Each dot represents a Bloch site (labeled in a picture for a given stack), and the electronic $\pi$ orbitals are highlighted with a given intensity based on the eigenvectors to our Hamiltonian. Note the edge state localization at sites $a_1$ and $b_4$. These are the low energy basis states in our Schrieffer-Wolf transform.}\label{fig:tetralayerABC}
\end{figure}

For ABC stacking, the origin of the flat band is due to edge state localization that causes our Hamiltonian to have a particularly low energy eigenstate solution. For an $N$-layer ABC stacking specifically, there exists only two sites in the Bloch bases that have no interlayer hopping; all other sites do. These are $a_1$ and $b_N$. Occupation of other sites would lead to higher energy eigenstates due to the energy associated with interlayer hopping, so $a_1$ and $b_N$ serve as the basis for the low energy domain for our Schrieffer-Wolf transformation. The low energy charge density distribution for tetralayer ABC stacking is plotted in Fig. ~\ref{fig:tetralayerABC}. Specifically, the eigenstates of our Hamiltonian give the amplitudes for which electrons may be localized at any given site, according to the tight binding model. There is clear localization on specifically the edge states, where the amount of hopping energy is minimal (there is no interlayer hopping). This minimizes energy, which explains the flat band.

\subsection{Multilayer Graphene with Arbitrary Stackings}

We now consider a multilayer graphene with a generalized stacking, which we call arbitrary stacking. Excluding twisting, arbitrary stacking can be represented by a string of the letters A, B, and C that will represent the relative orientation of each graphene layer. The cases AA, AB, and ABC stacking are special cases of this form of stacking.

Each well-understood stacking has its own properties \cite{Macdonald}. We may understand arbitrary stacking as an embedding of the stacking sequences for which we have studied earlier. The most studied cases are that of stacking faults, where we, in a sequence of either parallel, Bernal, or rhombohedral stacking, embed a small sequence of another one of the three common structures \cite{Koshino_nature}. For example, the sequence ABABABBABAB is one such fault, where we have embedded a parallel sequence within a Bernal sequence; the sequence ABCABCBCABCABC is another such fault, where we have embedded a Bernal sequence within a rhombohedral sequence. We will investigate the properties of the three stacking sequences and see how they evolve when we employ arbitrary stacking.

\begin{figure}[tbp]
\setkeys{Gin}{width=\linewidth}
\begin{subfigure}{.41\textwidth}
  \includegraphics{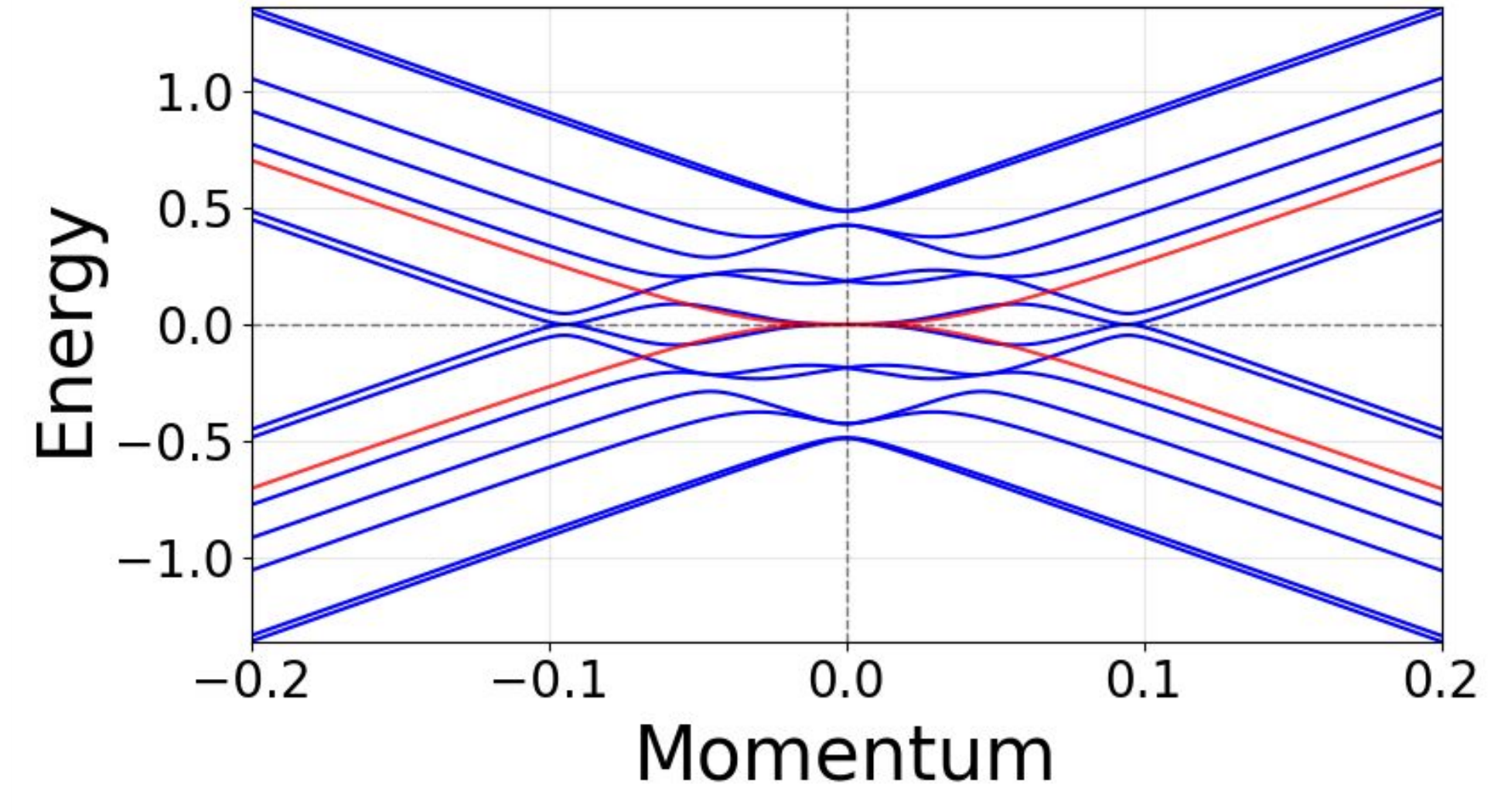}
  \caption{}
  \label{fig:4a}
\end{subfigure}
\hfill
\begin{subfigure}{.41\textwidth}
  \includegraphics{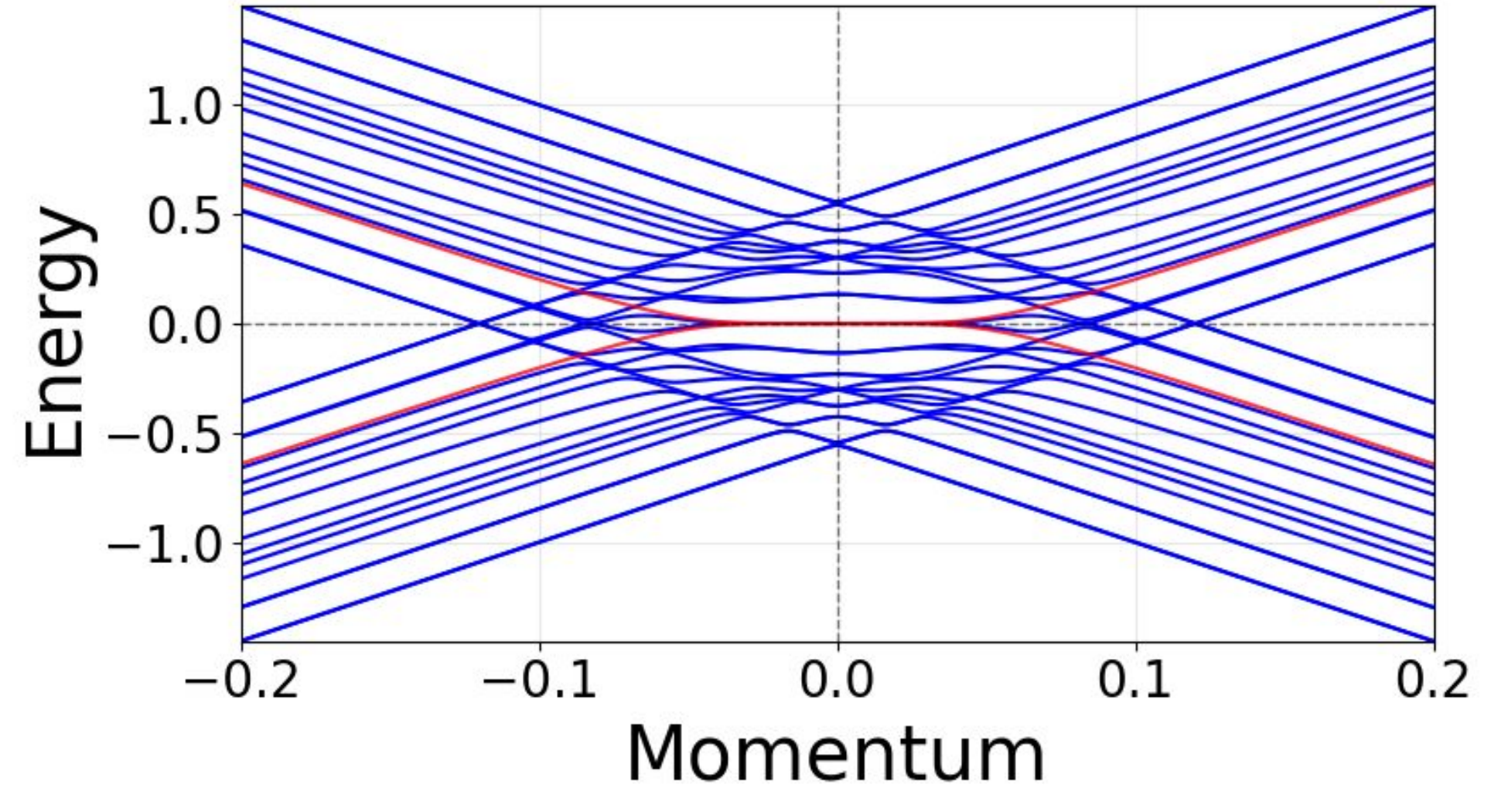}
  \caption{}
  \label{fig:4b}
\end{subfigure}

\begin{subfigure}{.41\textwidth}
  \includegraphics{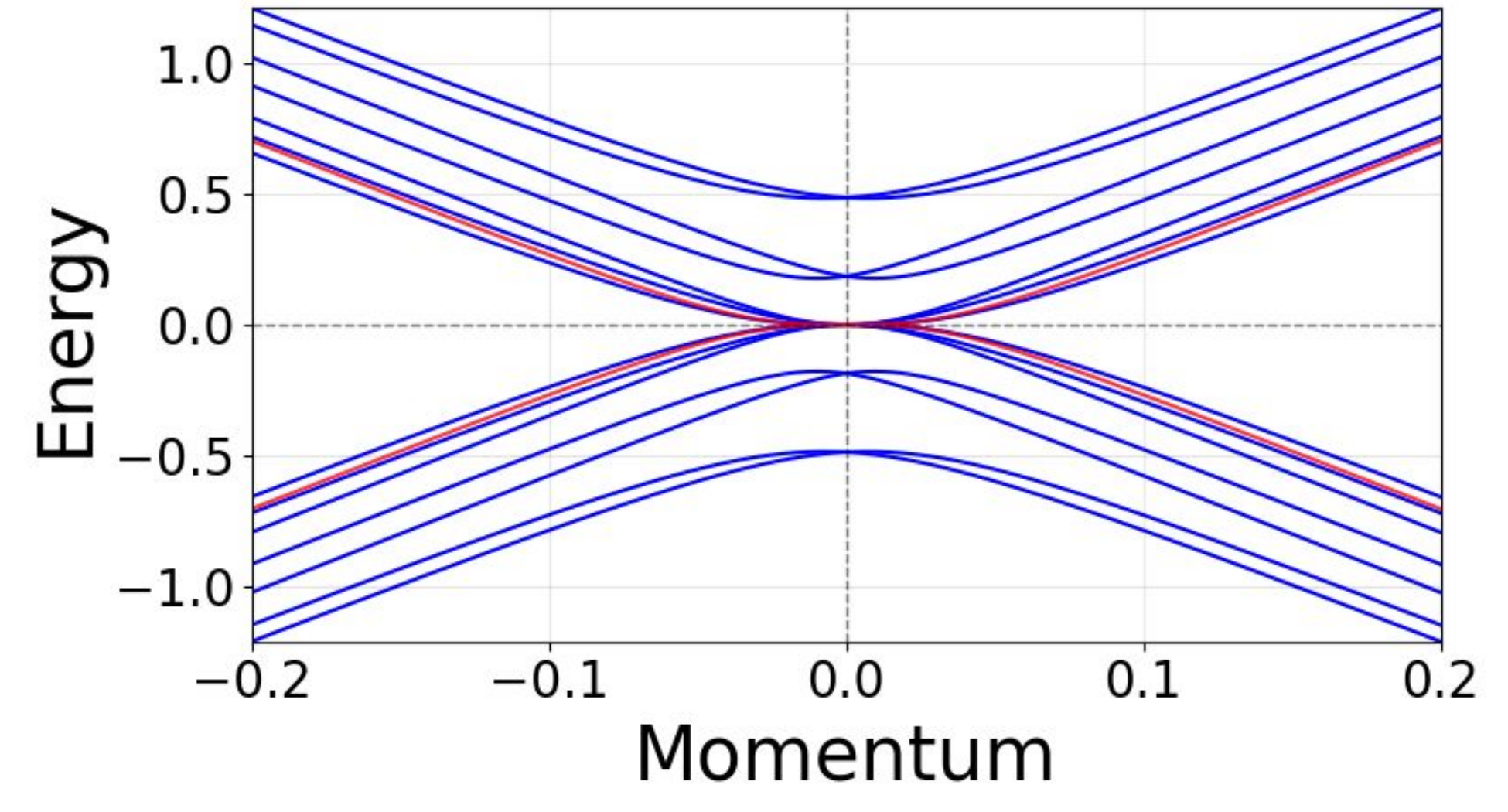}
  \caption{}
  \label{fig:4c}
\end{subfigure}
\hfill
\begin{subfigure}{0.41\textwidth}
    \includegraphics{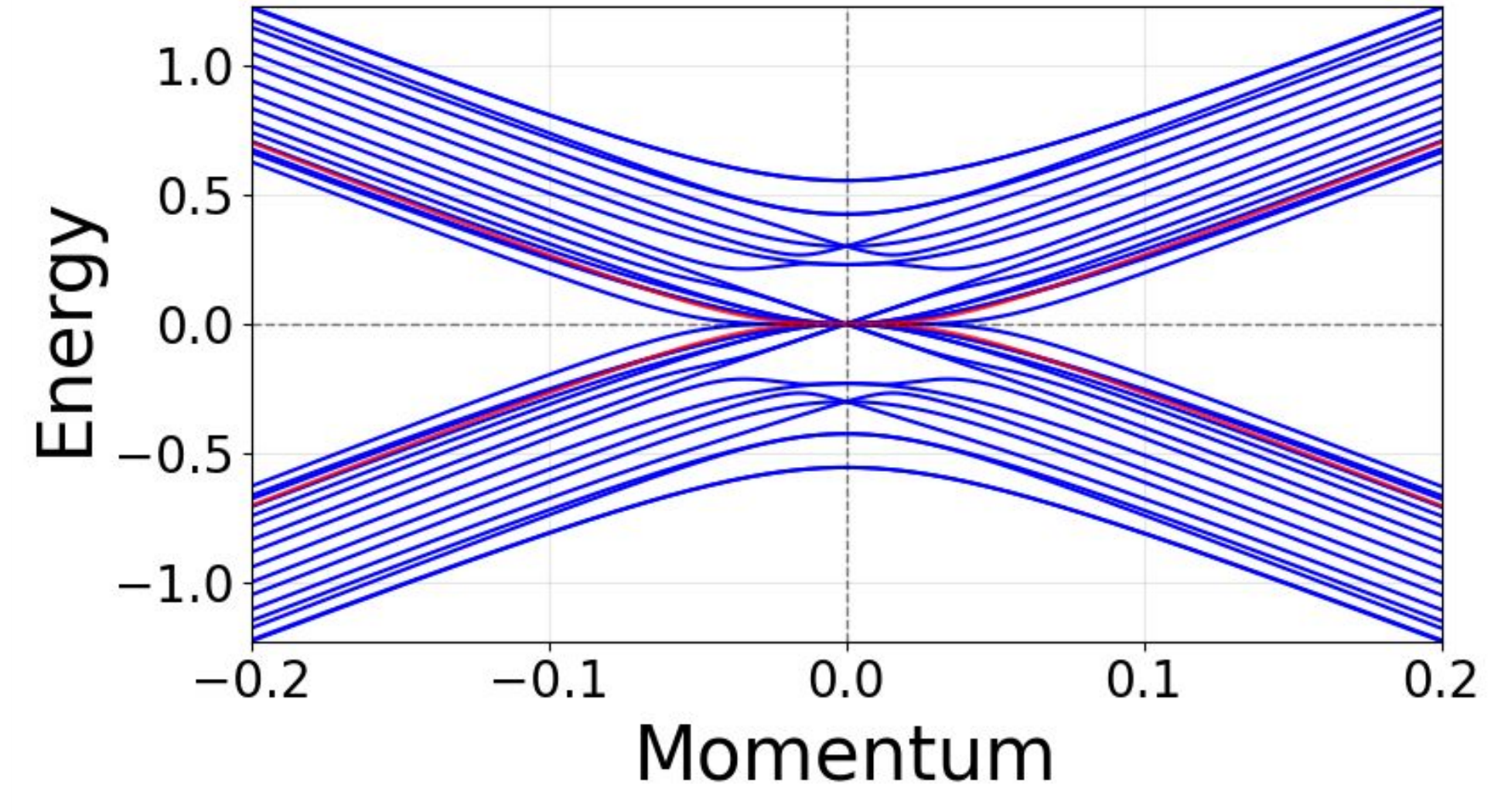}
    \caption{}
    \label{fig:4d}
\end{subfigure}
\captionsetup{justification=raggedright, singlelinecheck=false}
\caption{(a) Band dispersions for AAABCCC (blue). The dispersion for an ABC trilayer graphene is shown in red for comparison. Note how the flat band flattens the points of zero energy. (b) Band dispersions for AAAAAABCABCCCCCC (blue). The band dispersions for ABCABC are shown in red. (c) Band dispersions for ABABCBC (blue). The band dispersions for ABC trilayer graphene are shown in red. (d) Band dispersions for BABABABCABCBCBCB (blue) and for ABCABC (red). }
\label{fig:4}
\end{figure}

Firstly, we observe the emergence of the flat band. Fig.~\ref{fig:4} shows the dispersion relation for embedding ABC stacking into the parallel and Bernal stacking, respectively, along with the respective flat band for the embedded substacks; we see a striking resemblance between a flat band for our embedded ABC stacking and that of pure ABC stacking. It is thus possible to induce a flat band within parallel or Bernal stacking by inserting a rhombohedral stacking fault. This suggests the principle of decomposing a stack into components to analyze its properties, since the existence of ABC stacking within an arbitrary stack seems to be directly linked with the existence of a flat band.

\begin{figure}
  \setkeys{Gin}{width=\linewidth}
  \begin{subfigure}{.45\textwidth}
    \includegraphics{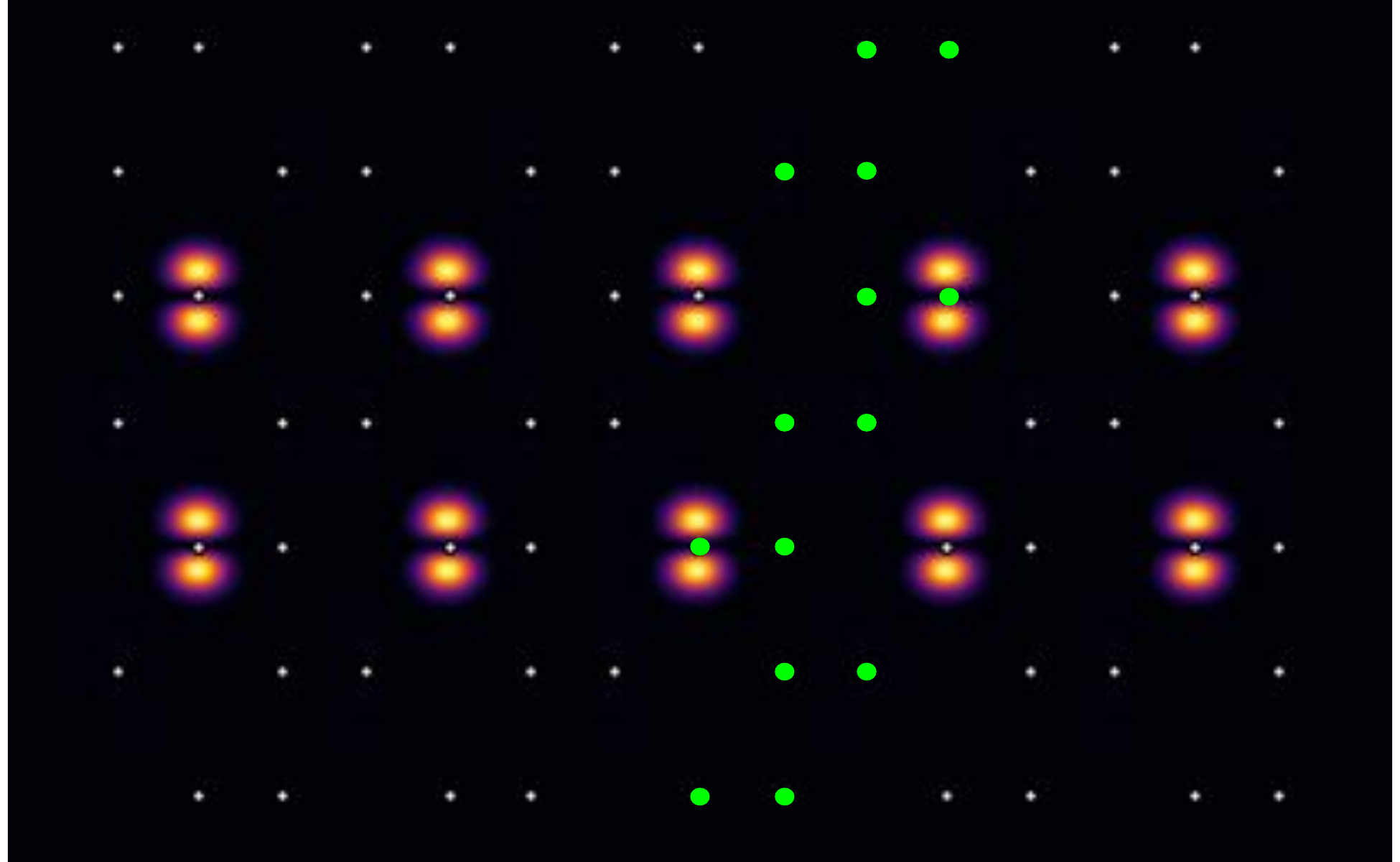}
    \caption{}
    \label{fig:5a}
  \end{subfigure}
  \hfill
  \begin{subfigure}{0.45\textwidth}
    \includegraphics{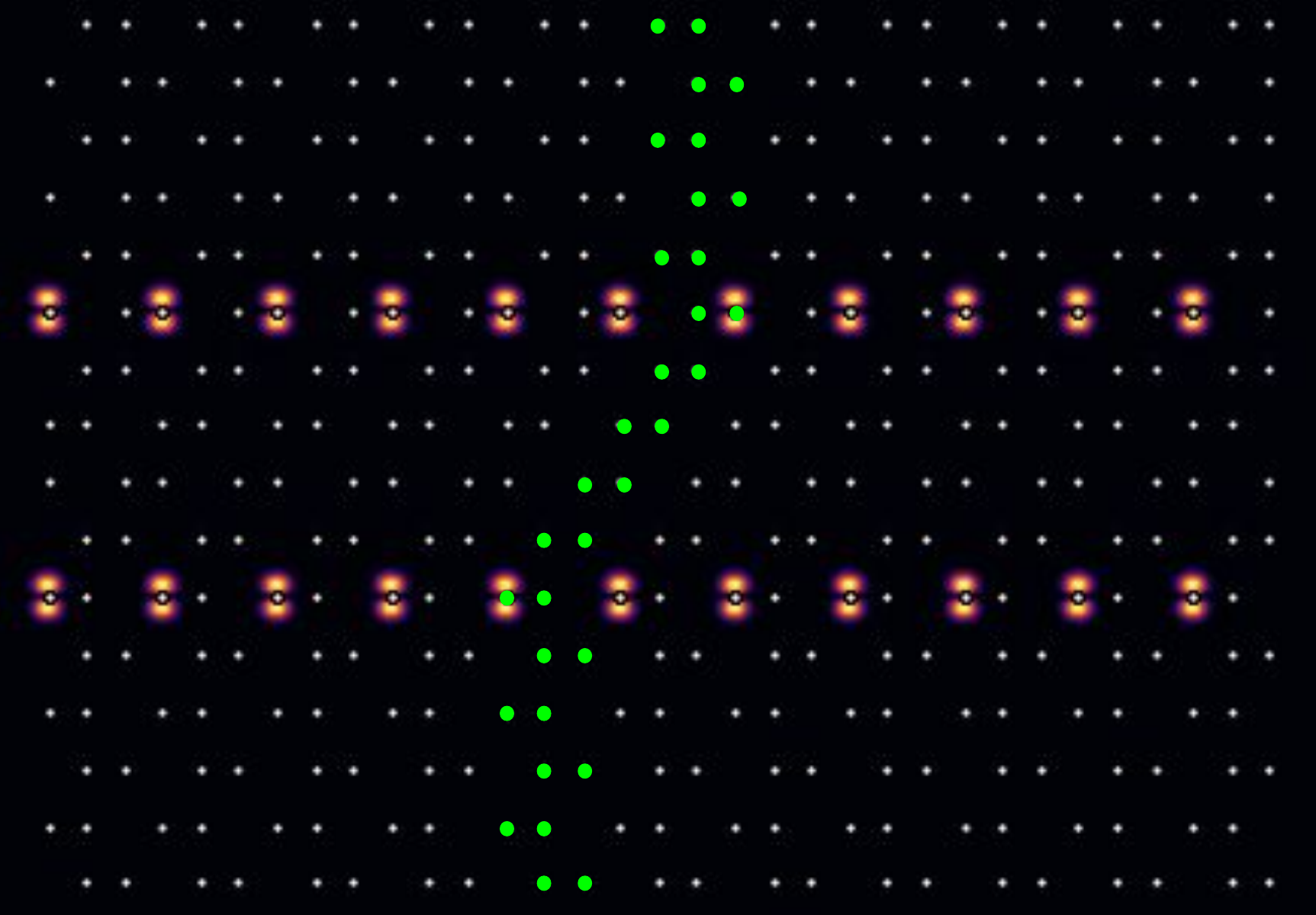}
    \caption{}
    \label{fig:5b}
  \end{subfigure}
  \captionsetup{justification=raggedright, singlelinecheck=false}
  \caption{Sufficiently low energy flat band eigenstate for (a) ABABCBC and (b) BABABABCABCBCBCB. For clarity, we have highlighted the stacking sequence of the carbon atoms using green dots. Note the localization at the edges of the embedded ABC substack. We do not see this for an ABC substack embedded within a parallel stacking.}
\end{figure}

Moreover, for embedding a rhombohedral substack into a Bernal stacking, it seems that we maintain the edge state localization within the ABC substack. This behavior is shown in Figs.~\ref{fig:5a} and ~\ref{fig:5b}, corresponding to the charge densities for the flat band exhibited by the dispersion relations plotted in Figs.~\ref{fig:4c} and ~\ref{fig:4d} respectively. For reference, we show in green the pattern followed by the stacking sequence. We have edge state localization, not at the edges of the stack, but instead at the edges of the ABC substack. We do not see the same localization when we embed an ABC substack into a parallel stacking. This is because, for AA stacking embedded with an ABC stack, the edges of the ABC substack no longer preserves the isolated sites with only intralayer coupling. Instead, we are forced to have interlayer coupling for the ABC edge sites, which no longer makes them necessarily low energy. For ABC embedded Bernal stacking, however, the edge states still preserve their low energy property.

One notable property of the AB stacking with an ABC fault is the presence of both a linear and a flat band. As we have already shown, a flat band can be induced by embedding an ABC stack into any arbitrary graphene stack. Moreover, Eq.~\ref{eq:AB_disp} can exhibit linear bands when $N$ is odd, allowing $r=\frac{N+1}{2}$ to be an integer so that $t_\perp \cos(\frac{r\pi}{N+1}) = 0$. This creates a linear band $\epsilon^{\pm,AB}_{\frac{N+1}{2},\vec{p}}=\pm v|\vec{p}|$, that is exactly the same as Eq.~\ref{eq:massless_dirac_fermions}. It is understood that flat bands, on the scale induced by large ABC stacks, are indicative of strongly correlated electronic systems. Furthermore, the linear band of Eq.~\ref{eq:massless_dirac_fermions} represent the existence of massless Dirac fermions. For certain AB stacking embedded with ABC stacking, the possibility to simultaneously maintain a linear band and induce a flat band implies the coexistence of strongly correlated electrons with massless Dirac fermions. Thus, these stacks could be viable in providing insight to the interactions of these particles.

Figs.~\ref{fig:4a} and ~\ref{fig:4b} show the dispersion relations for AA stacking with an ABC fault. In AA stacking, we have multiple zero-energy points that are understood as rings about the Dirac point. Electrons will fill these regions first before moving upwards in energy \cite{Aufbau}. Figs.~\ref{fig:4c} and ~\ref{fig:4d} show the dispersion relations for AB stacking with an ABC fault. In AB stacking, the zero-energy locations occur only at the Dirac point. We may induce this property of AA stacking into non-parallel stacking by means of stacking faults. Notice that the stacks that include some parallel stacking are exclusively the ones that have band crossings at nonzero $|\vec{p}|$ values from the $\vec{K}$, namely Figs.~\ref{fig:4a} and ~\ref{fig:4b}. In fact, we can make an analytical statement about the direct linkage between AA stacking and these nontrivial band crossings, moreover finding the exact $\vec{p}$ for which these occur.
Zero eigenvalues corresponds to our stacking matrix
\begin{equation}
  H(\vec{p}) = t_\perp\begin{pmatrix}
  0 & f\pi^{\dagger} & \beta_1 & \gamma_1 & 0 & 0 & \cdots & 0 & 0 \\
  f\pi & 0 & \alpha_1 & \beta_1 & 0 & 0 & \cdots & 0 & 0 \\
  \beta_1 & \alpha_1 & 0 & f\pi^{\dagger} & \beta_2 & \gamma_2 & \cdots & 0 & 0 \\ \gamma_1 & \beta_1 & f\pi & 0 & \alpha_2 & \beta_2 & \cdots & 0 & 0 \\
  0 & 0 & \beta_2 & \alpha_2 & 0 & f\pi^{\dagger} & \cdots & 0 & 0 \\
  0 & 0 & \gamma_2 & \beta_2 & f\pi & 0 & \cdots & 0 & 0 \\
  \vdots & \vdots & \vdots & \vdots & \vdots & \vdots & \ddots & \vdots & \vdots \\
  0 & 0 & 0 & 0 & 0 & 0 & \cdots & 0 & f\pi^{\dagger} \\
  0 & 0 & 0 & 0 & 0 & 0 & \cdots & f\pi & 0
\end{pmatrix}
\end{equation}
having zero determinant. Then define $\chi_i$ for $i \in \{1,2,3,...2N\}$ to be the determinant from rows and columns $i$ to $2N$, or in other words the determinant of the bottom right submatrix of dimensions $(2N-i+1)\times(2N-i+1)$ (working from left to right and top to bottom, this submatrix would be constructed by taking the $i$th row and column and everything onward). We hence establish by definition
\begin{equation}
    \chi_{2N-1} = -v^2|\vec{p}|^2.
\end{equation}
Then, let us work for $k \in \{1,2,3,...N-1\}$. Then, $\chi_{2k-1}$ may be interpreted as the determinant of a substack of $N-k+1$ layers, starting from the bottom of our original stack of $N$ layers. If $\alpha_k = 1$, we have
\begin{equation}
    \chi_{2k-1} = -v^2|\vec{p}|^2\chi_{2(k+1)-1} \label{eq:alpha_recur},
\end{equation}
where we use the fact that if a matrix has an entire column of zeroes, its determinant is zero. If $\gamma_k = 1$, we have similarly have
\begin{equation}
    \chi_{2k-1} = -v^2|\vec{p}|^2\chi_{2(k+1)-1} \label{eq:gamma_recur}.
\end{equation}

We see that, for a stack whose top two sheets are connected by an AC or AB connection (so that there is a shift between them), the determinant of the original substack, denoted by $\chi_{2k-1}$, is $0$ for $|\vec{p}| \neq 0$ if and only if its determinant taking the topmost sheet off, denoted by $\chi_{2(k+1)-1}$, is also $0$. Then, having the determinant be $0$ is impossible if we only have Bernal and rhombohedral stacking as the recurrence relation has shown (the base case, for monolayer graphene, is always nonzero for $|\vec{p}| \neq 0$). Essentially, we have shown that the determinant of a stack of size $N$ represented by a string where adjacent characters are different (ABABABC is one such string, ABBAB is not) is equal to
\begin{equation}
    \det(H) = (-1)^Nv^{2N}|\vec{p}|^{2N},
\end{equation}
which is 0 if and only if $|\vec{p}| = 0$. AA stacking does not follow such a simple recurrence relation. Hence, it is possible for it to induce multiple Dirac points. For a substack that is on top an AA pattern, which would be the $\beta_k = 1$ case, we can write in block matrix form:
\begin{equation}
    H = \begin{pmatrix}\textbf{M} & t_\perp\textbf{J} \\ t_\perp\textbf{J}^T & \textbf{S}\end{pmatrix},
\end{equation}
where \textbf{M} is the Hamiltonian for monolayer graphene, \textbf{S} is the Hamiltonian for the substack if we take the top layer (connected by an AA stacking connection) off. $\textbf{J}$ is defined as
\begin{equation}
    \textbf{J} = \begin{pmatrix}1 & 0 & 0 & \cdots & 0\\ 0 & 1 & 0 & \cdots & 0\end{pmatrix}
\end{equation}
to keep the matrix square. For such a block matrix, we may use the determinant formula \cite{invertable_det}
\begin{equation}
    \det(H) = \det(\textbf{M})\det(\textbf{S} - t_\perp^2\textbf{J}^T\textbf{M}^{-1}\textbf{J}) .\label{eq:det_formula}
\end{equation}
which implies that $\det(H) = 0$ if and only if $\det(\textbf{S} - t_\perp^2\textbf{J}^T\textbf{M}^{-1}\textbf{J}) = 0$.

Of course,
\begin{equation}
    \textbf{M}^{-1} = \frac{1}{v^2|\vec{p}|^2}\begin{pmatrix} 0 & v\pi^\dagger \\ v\pi & 0\end{pmatrix},
\end{equation}
meaning that
\begin{equation}
    t_\perp^2\textbf{J}^T\textbf{M}^{-1}\textbf{J} = \frac{t_\perp^2}{v^2|\vec{p}|^2}\begin{pmatrix}
0 & v\pi^\dagger & 0 & 0 & \cdots & 0 \\
v\pi & 0 & 0 & 0 & \cdots & 0 \\
0 & 0 & 0 & 0 & \cdots & 0 \\
0 & 0 & 0 & 0 & \cdots & 0 \\
\vdots & \vdots & \vdots & \vdots & \ddots & \vdots \\
0 & 0 & 0 & 0 & \cdots & 0
\end{pmatrix}.
\end{equation}

If we now assume the next sequence has $\beta_{k-1} = 0$, then $t_\perp^2\textbf{J}^T\textbf{M}^{-1}\textbf{J}$ can be written in the matrix form:
\begin{equation}
    \resizebox{0.47\textwidth}{!}{$\textbf{S} - t_\perp^2\textbf{J}^T\textbf{M}^{-1}\textbf{J} = \begin{pmatrix}
0 & (1-\frac{t_\perp^2}{v^2|\vec{p}|^2})v\pi^\dagger & 0 & \gamma_{k+1}t_\perp & \cdots & 0 \\
(1-\frac{t_\perp^2}{v^2|\vec{p}|^2})v\pi & 0 & \alpha_{k+1}t_\perp & 0 & \cdots & 0 \\
0 & \alpha_{k+1}t_\perp & 0 & v\pi^\dagger & \cdots & 0 \\
\gamma_{k+1}t_\perp & 0 & v\pi & 0 & \cdots & 0 \\
\vdots & \vdots & \vdots & \vdots & \ddots & \vdots \\
0 & 0 & 0 & 0 & \cdots & 0 \end{pmatrix}$},
\end{equation}
whose determinant we gather from the same method we used to derive \ref{eq:alpha_recur} and \ref{eq:gamma_recur} to be
\begin{equation}
    \det(\textbf{S} - t_\perp^2\textbf{J}^T\textbf{M}^{-1}\textbf{J}) = -(1-\frac{t_\perp^2}{v^2|\vec{p}|^2})^2v^2|\vec{p}|^2\chi_{2(k+2)-1}.
\end{equation}

This generates a zero determinant at $|\vec{p}| \neq 0$, namely $|\vec{p}| = \frac{t_\perp}{v}$. We may also generalize this to $L$ parallel connections (which is to say we embed a length $L+1 = N$ parallel stacks). In general, this determinant is 0 if we have a $2\times2$ block of zeros in the top left. Define
\begin{equation}
    Q_0(\vec{p}) = 0 \label{eq:Q_base}
\end{equation}
\begin{equation}
    Q_{n+1}(\vec{p}) = \frac{t_\perp^2}{(1-Q_n(\vec{p}))v^2|\vec{p}|^2}.
\end{equation}

We would like to show inductively that we can solve for $|\vec{p}| \neq 0$ that gives zero determinant by setting
\begin{equation}
    Q_L(\vec{p}) = 1 \label{eq:Q_eq}.
\end{equation}
As we discussed earlier, this only happens if in the place of top left monolayer of the resultant difference is a matrix of zeros. In general, we would have our matrix whose determinant we set to 0 to be
\begin{equation}
    \resizebox{0.47\textwidth}{!}{$\begin{pmatrix}\textbf{M} & t_\perp\textbf{J} \\ t_\perp\textbf{J}^T & \textbf{S}\end{pmatrix} = \begin{pmatrix}
0 & (1-Q_n(\vec{p}))v\pi^\dagger & \beta t_\perp & \gamma t_\perp & \cdots & 0 \\
(1-Q_n(\vec{p}))v\pi & 0 & \alpha t_\perp& \beta t_\perp& \cdots & 0 \\
\beta t_\perp& \alpha t_\perp& 0 & v\pi^\dagger & \cdots & 0 \\
\gamma t_\perp& \beta t_\perp& v\pi & 0 & \cdots & 0 \\
\vdots & \vdots & \vdots & \vdots & \ddots & \vdots \\
0 & 0 & 0 & 0 & \cdots & 0 \end{pmatrix}$},
\end{equation}
since our inductive step requires us to assume this is true. More concretely, we are requiring that the above matrix has zero determinant for the $\beta =0$ case, at the end of the $N$ layer stack where $L=N-1$, which implies Eq. \eqref{eq:Q_eq}. Essentially, our mathematical formula that creates a subproblem to zero the determinant of $\textbf{S} - t_\perp^2\textbf{J}^T\textbf{M}^{-1}\textbf{J}$ essentially induces an effective momentum elements $v\pi \mapsto (1-Q_n(\vec{p}))v\pi$ which we need to zero at $\alpha=1$ or $\gamma = 1$ connections. To proceed with induction, note we have already solved a base case for for we have only one parallel connection. We take the $\beta = 1$ case and we solve with the formula given in Eq. \eqref{eq:det_formula}, noting that
\begin{equation}
    \resizebox{0.45\textwidth}{!}{$\textbf{M}^{-1} = \frac{1}{((1-Q_n(\vec{p}))v)^2|\vec{p}|^2}\begin{pmatrix}0 & (1-Q_n(\vec{p}))v\pi^\dagger \\ (1-Q_n(\vec{p}))v\pi & 0 \end{pmatrix}$}.
\end{equation}
Calculating $\textbf{S} - t_\perp^2\textbf{J}^T\textbf{M}^{-1}\textbf{J}$, we therefore need to zero the determinant of
\begin{equation}
    \resizebox{0.45\textwidth}{!}{$\begin{pmatrix}
0 & (1-\frac{t_\perp^2}{(1-Q_n(\vec{p}))v^2|\vec{p}|^2})v\pi^\dagger & \beta_- t_\perp& \gamma_- t_\perp& \cdots & 0 \\
(1-\frac{t_\perp^2}{(1-Q_n(\vec{p}))v^2|\vec{p}|^2})v\pi & 0 & \alpha_- t_\perp& \beta_- t_\perp& \cdots & 0 \\
\beta_- t_\perp& \alpha_- t_\perp& 0 & v\pi^\dagger & \cdots & 0 \\
\gamma_- t_\perp& \beta_- t_\perp& v\pi & 0 & \cdots & 0 \\
\vdots & \vdots & \vdots & \vdots & \ddots & \vdots \\
0 & 0 & 0 & 0 & \cdots & 0 \end{pmatrix}$},
\end{equation}
where $(\alpha_-, \beta_-, \gamma_-)$ precedes $(\alpha,\beta,\gamma)$. This is equal to by definition
\begin{equation}
    \resizebox{0.45\textwidth}{!}{$\begin{pmatrix}
0 & (1-Q_{n+1}(\vec{p}))v\pi^\dagger & \beta_- t_\perp& \gamma_- t_\perp& \cdots & 0 \\
(1-Q_{n+1}(\vec{p}))v\pi & 0 & \alpha_- t_\perp& \beta_- t_\perp& \cdots & 0 \\
\beta_- t_\perp& \alpha_- t_\perp& 0 & v\pi^\dagger & \cdots & 0 \\
\gamma_- t_\perp& \beta_- t_\perp& v\pi & 0 & \cdots & 0 \\
\vdots & \vdots & \vdots & \vdots & \ddots & \vdots \\
0 & 0 & 0 & 0 & \cdots & 0 \end{pmatrix}$},
\end{equation}
which thus completes the inductive step. To solve this recurrence, we define $x = \frac{v|\vec{p}|}{t_\perp}$, and $f_n(x) = 1-Q_n(\vec{p})$, which gives us
\begin{equation}
    f_{n+1}(x) = 1-\frac{1}{f_n(x)x^2}.
\end{equation}
Our goal now is to find the roots of $f_L(x)$. We note our recurrence can be written as
\begin{equation}
    xf_{n+1}(x) = x-\frac{1}{f_n(x)x}.
\end{equation}
We then consider Chebyshev polynomials of the second kind \cite{Chebyshev}:
\begin{equation}
    U_{n+1}(y) = 2yU_n(y) - U_{n-1}(y),
\end{equation}
and note this can be written for $y=\frac{x}{2}$ as
\begin{equation}
    \frac{U_{n+1}(\frac{x}{2})}{U_n(\frac{x}{2})} = x - \frac{U_{n-1}(\frac{x}{2})}{U_n(\frac{x}{2})}.
\end{equation}

We would like to show $xf_n(x) = \frac{U_{n+1}(\frac{x}{2})}{U_n(\frac{x}{2})}$. Notice that the recurrence relations are automatically identical. We know $f_0(x) = 1$ from Eq. \eqref{eq:Q_base}. Since $U_{1}(x) = 2x$ and $U_0(x) = 1$, we have $\frac{U_{1}(\frac{x}{2})}{U_0(\frac{x}{2})} = x = xf_0(x)$, so we have verified both the base case and the recurrence relation, which completes the proof. The roots $x_r$ that satisfy $U_D(x_r) = 0$ are given by
\begin{equation}
    x_r = \cos(\frac{r\pi}{D+1})
\end{equation}
for $r \in \{1,2,3,...D\}$. Then we may substitute $x = \frac{x_r}{2}$ and $|\vec{p}| = \frac{t_\perp x}{v}$ with $D = L+1$ so that for $\vec{p}$ to satisfy Eq. \eqref{eq:Q_eq}, it is necessary that
\begin{equation}
    |\vec{p}| = 2\frac{t_\perp}{v}\cos(\frac{r\pi}{L+2}).
\end{equation}
Since $L$ is the number of connections, it may be more useful to consider the number of layers $N = L+1$. We have
\begin{equation}
    |\vec{p}| = 2\frac{t_\perp}{v}\cos(\frac{r\pi}{N+1}) \label{eq:all_zeros}.
\end{equation}

Technically, we would also have to exclude the roots of $U_L$, but they will in fact never show up in our expression, because $\gcd(n,n+1) = 1$ for $n \in \mathbb{Z}^+$. More importantly, we must carefully consider $|\vec{p}| = 0$. We have implicitly assumed this not to be true when we inverted our monolayer graphene block matrix; if $|\vec{p}| = 0$, then our monolayer graphene matrix becomes singular. Moreover, since $xf_n(x) = \frac{U_{n+1}(\frac{x}{2})}{U_n(\frac{x}{2})}$, we'd have to exclude $x=0 \implies |\vec{p}| = 0$ as a necessary root of $f_N(x)$ anyway. However, Eq. \eqref{eq:all_zeros} gives $|\vec{p}| = 0$ for odd N. We can do an inductive proof to show that indeed $0$ momentum gives $0$ energy for odd N only. Our strategy for proof will be to try to find nonzero determinant with matrix minors, and then show that we can do so for odd N but not even N. Define the substack, taking off all the layers until we leave the complete AA embed on top to be \textbf{S}. We will now stack an AA bilayer on top
\begin{equation}
    H = \begin{pmatrix}\textbf{M$_2$} & \textbf{t} \\ \textbf{t}^T & \textbf{S}_2\end{pmatrix},
\end{equation}
where
\begin{equation}
    \textbf{M$_2$} = \begin{pmatrix}
0 & 0 & t_\perp & 0\\
0 & 0 & 0 & t_\perp \\
t_\perp & 0 & 0 & 0 \\
0 & t_\perp & 0 & 0\end{pmatrix}
\end{equation}
and
\begin{equation}
    \textbf{t} = \begin{pmatrix}0 & 0 & 0 & \cdots &0 \\ 0 & 0 & 0 &\cdots & 0 \\ t_\perp & 0 &0 &\cdots & 0\\ 0 & t_\perp & 0 &\cdots & 0\end{pmatrix},
\end{equation}
again, to keep the matrix square. We realize
\begin{equation}
    \det(H) = t_\perp^4\det(\textbf{S}_2), \label{eq:ifsingular}
\end{equation}
by using the cofactor expansion to calculate the determinant of our Hamiltonian. Essentially, the only possibly nonzero matrix minor is the one that come from the $t_\perp$ elements in $\textbf{M}_2$. If we took a $t_\perp$ element from $\textbf{t}_2$ instead of $\textbf{M}_2$, the resultant submatrix would have a column of zeroes, which would mean a matrix minor of zero. In doing this, the only nonzero matrix minor would be the determinant of $\textbf{S}_2$. It would then be trivial to consider $N = 0$ and $N= 1$. Furthermore, if at any point in our recurrence relation $f_n(x) = 0$ for $n<N$, we would have $\textbf{M}$ become singular (specifically, a $2\times2$ matrix of zeros), making Eq.~\ref{eq:det_formula} invalid. To resolve this, consider that
\begin{equation}
    \textbf{M$_2$} = \begin{pmatrix}
0 & 0 & t_\perp & 0\\
0 & 0 & 0 & t_\perp \\
t_\perp & 0 & 0 & v\pi^\dagger \\
0 & t_\perp & v\pi & 0\end{pmatrix}.
\end{equation}
By similar arguments, Eq.~\ref{eq:ifsingular} holds. This means that the effect in general of having a singular $\textbf{M}$ at any step will be that the determinant of our Hamiltonian is zero if the determinant $\textbf{S}_2$ is zero. Notice that if $f_n(x) = 0$, then $f_{n+2}(x) = 1$, which means that our recurrence relation is consistent with this idea and therefore is valid in both sufficiency and completeness. Hence, it is in fact true that every single solution of Eq. \ref{eq:all_zeros} conveniently gives zero energy, including the $|\vec{p}| = 0$ solutions. What we have shown is that the zero energy momenta due to an embedded N-layer AA stack are the exact same momentum for zero energy for the N-layer AA stack itself; Eq. \eqref{eq:all_zeros} and Eq. \eqref{eq:zero_AA_eigen} are identical for zero-energy eigenvalues (one could, if preferred, also include absolute value signs in Eq. \eqref{eq:all_zeros} since there is positive and negative symmetry of cosine around $\frac{\pi}{2}$). Moreover, due to the recurrence relations in Eq. \eqref{eq:alpha_recur} and \eqref{eq:gamma_recur}, Eq. \eqref{eq:all_zeros} in fact gives every single zero-energy momentum.


Our analysis of completely arbitrary stacking is simply a generalization of stacking faults. It is understood by its derivation that Eq. \eqref{eq:all_zeros} is true for all arbitrary stacks. In general, our graphene stack is represented by a string of A, B, and C. And every stack can be represented by the stacking of parallel substacks. If we denote some stack to be $S$, then there exists a sequence $S_i$ for $i \in \{1,2,3,...K\}$ such that
\begin{equation}
    S = \bigoplus_{i=1}^PS_i
\end{equation}
and we guarantee that each $S_i$ is a string of the same letter. Our notation is simply for string concatenation. For example, if $S_1$ is an AA stack and $S_2$ is a BB stack, then $S = \bigoplus_{i=1}^2S_i$ is an AABB stack. Moreover, we establish by necessity that for $j \in \{1,2,3...K-1\}$, $S_j$ and $S_{j+1}$ are of different letters. Then denote $N_i$ to be the lengths of each parallel stack. For example, BABBBCCCC is composed of B, A, BBB, and CCCC, so in this case $N_i \in \{1,3,4\}$. Note how $N_i$ for the stack BABBBCCCC does not contain the element $2$. Now, denote $P_i = \{p_{1i}, p_{2i}, p_{3i}... p_{Ni}\}$ so that
\begin{equation}
    p_{r_ii} = 2\frac{t_\perp}{v}\cos(\frac{r_i\pi}{N_i+1}) \label{eq:all_p_value}
\end{equation}
for $r_i \in \{1,2,3...N_i\}$. Finally, define
\begin{equation}
    P_0 = \bigcup_{i=1}^KP_i.
\end{equation}
Then, $\vec{p}$ gives a zero energy eigenvalue for our Hamiltonian $H_S(\vec{p})$ if and only if $|\vec{p}| \in P_0$. In other words, if such a $\vec{p}$ gives zero energy for any one of the parallel stacked components $S_i$ of $S$ then it gives zero energy for $S$. Eq. \ref{eq:all_p_value} therefore suggests that the properties of arbitrarily stacked mulitlayer graphene are determined by considering its substacks in isolation. We have done so by analyzing the role of parallel stacking in an isolated stack. Without parallel stacking, such a phenomenon may also occur in an arbitrary stack consisting of only Bernal and rhombohedral stacking \cite{Koshino_2013}. It has been shown that band structure analysis for low momentum in this case may be done by focusing on the Bernal components in isolation. Hence, the behavior follows our idea above. The electronic properties of arbitrarily stacked multilayer graphene therefore in general appear to be adopted from specific subsequences of its stacking order.

\begin{figure*}
  \setkeys{Gin}{width=\linewidth}
  \begin{subfigure}{0.33\textwidth}
    \includegraphics{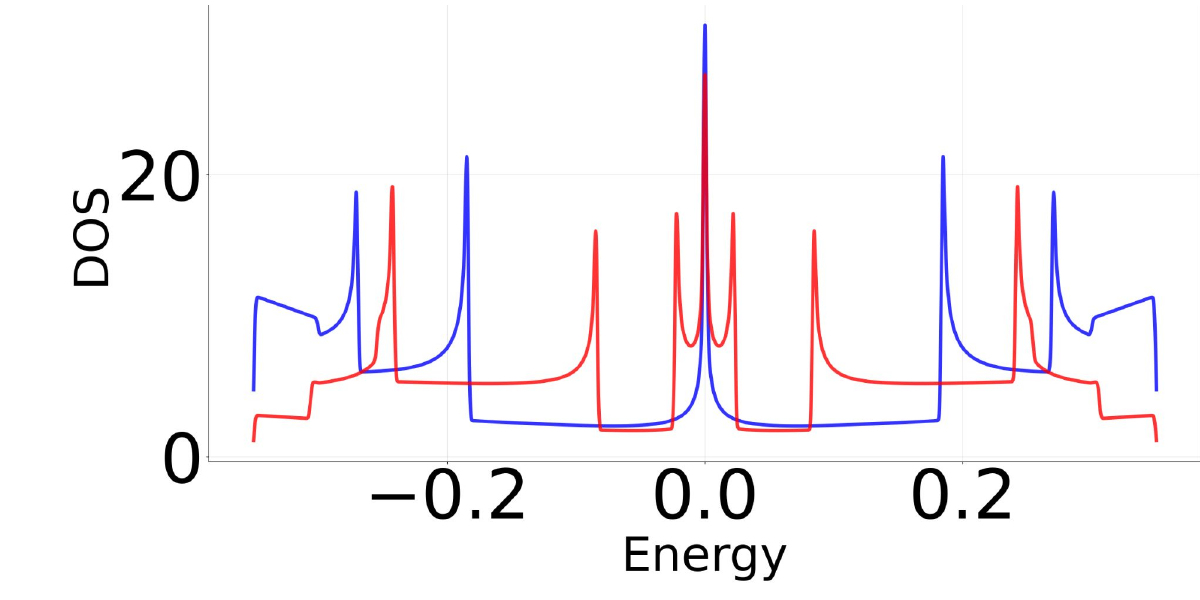}
    \caption{}
    \label{fig:6a}
  \end{subfigure}
  \begin{subfigure}{0.32\textwidth}
    \includegraphics{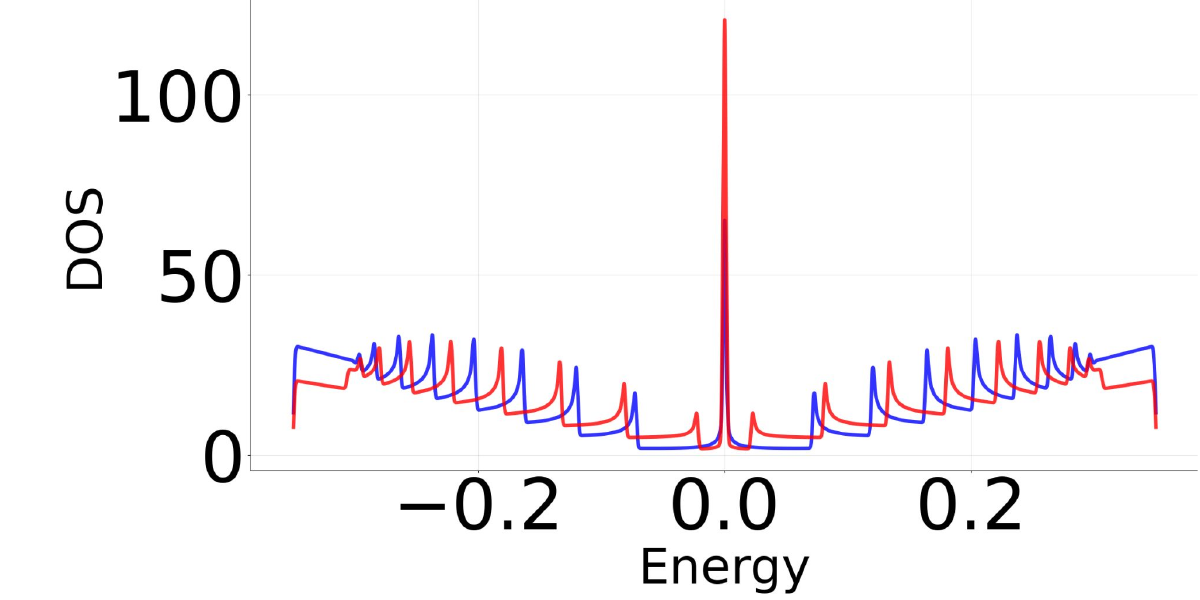}
    \caption{}
    \label{fig:6b}
  \end{subfigure}
  \begin{subfigure}{0.33\textwidth}
    \includegraphics{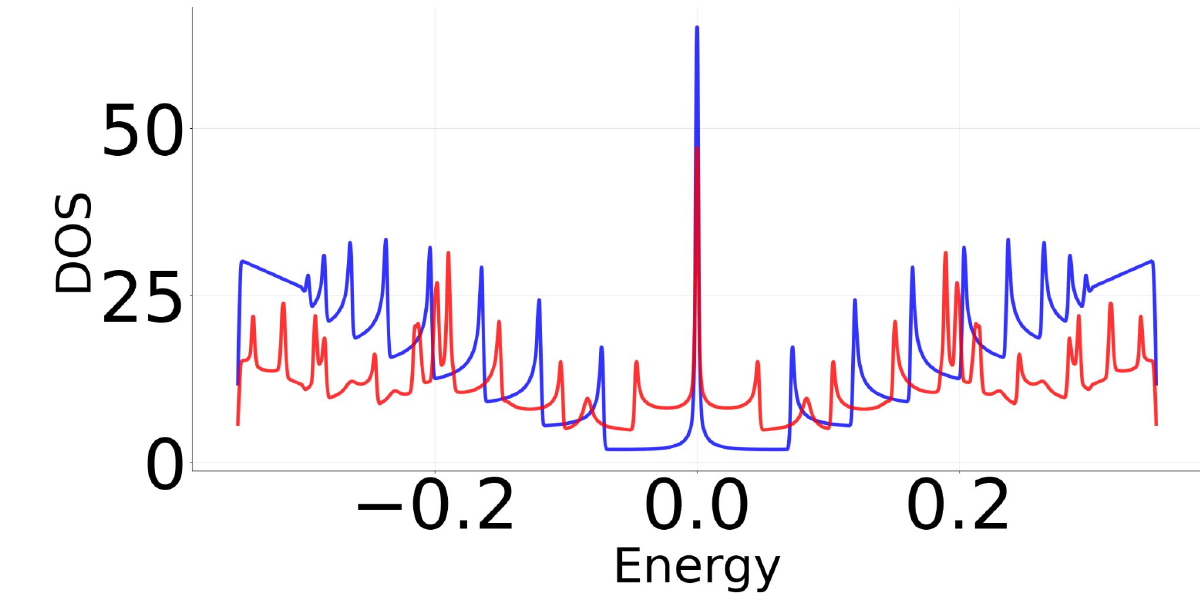}
    \caption{}
    \label{fig:6c}
  \end{subfigure}
  \hfill
  \begin{subfigure}{0.33\textwidth}
    \includegraphics{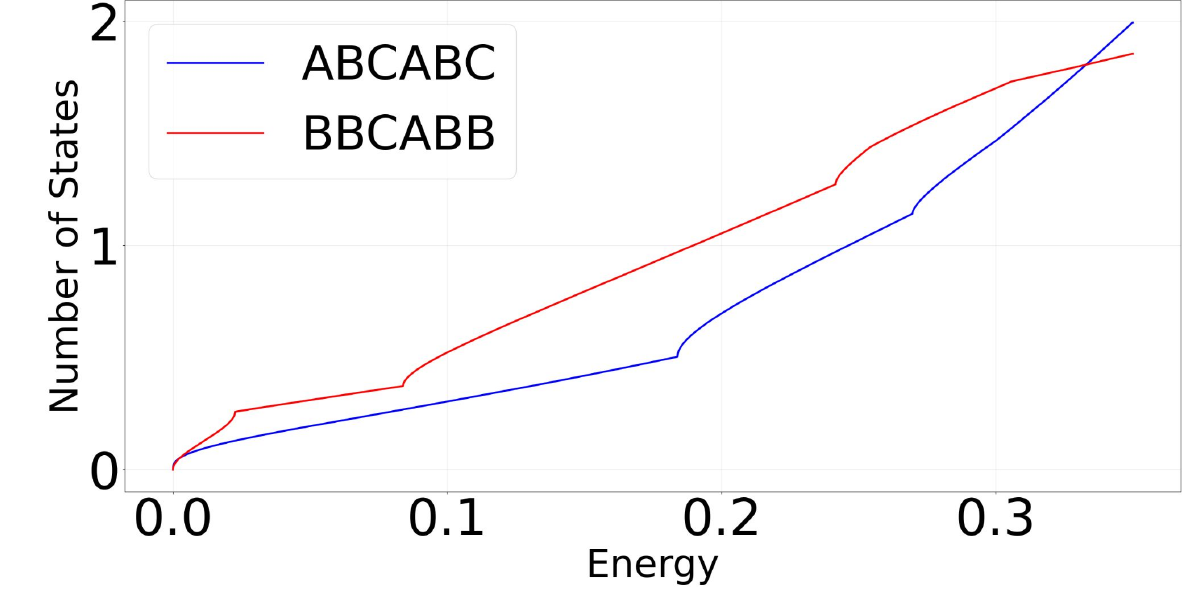}
    \caption{}
    \label{fig:6d}
  \end{subfigure}
  \begin{subfigure}{0.32\textwidth}
    \includegraphics{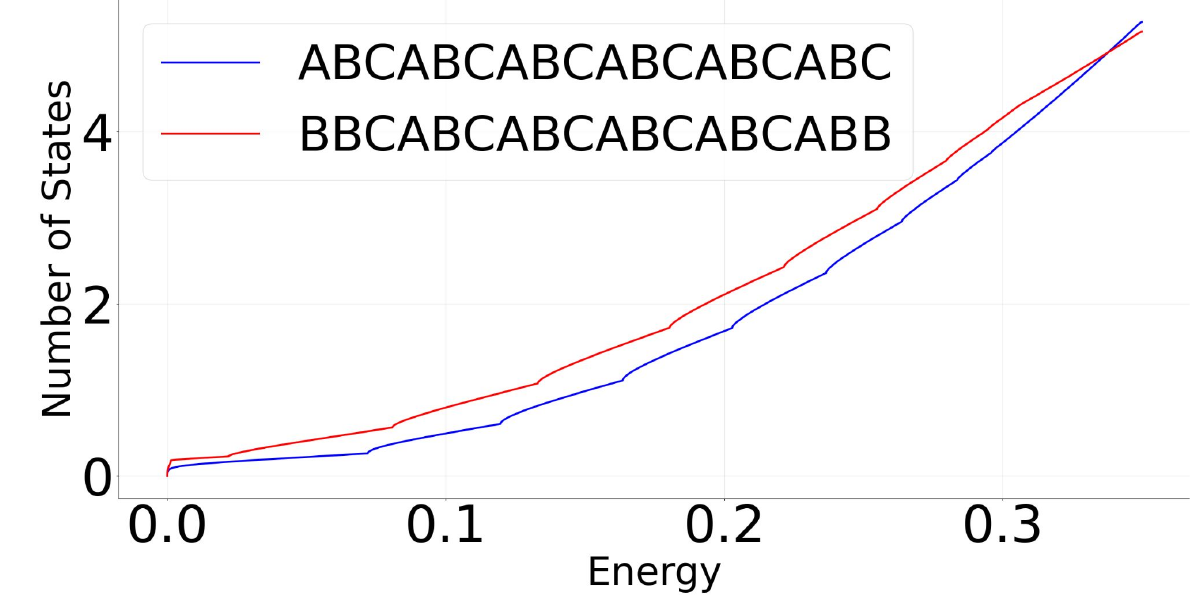}
    \caption{}
    \label{fig:6e}
  \end{subfigure}
  \begin{subfigure}{0.33\textwidth}
    \includegraphics{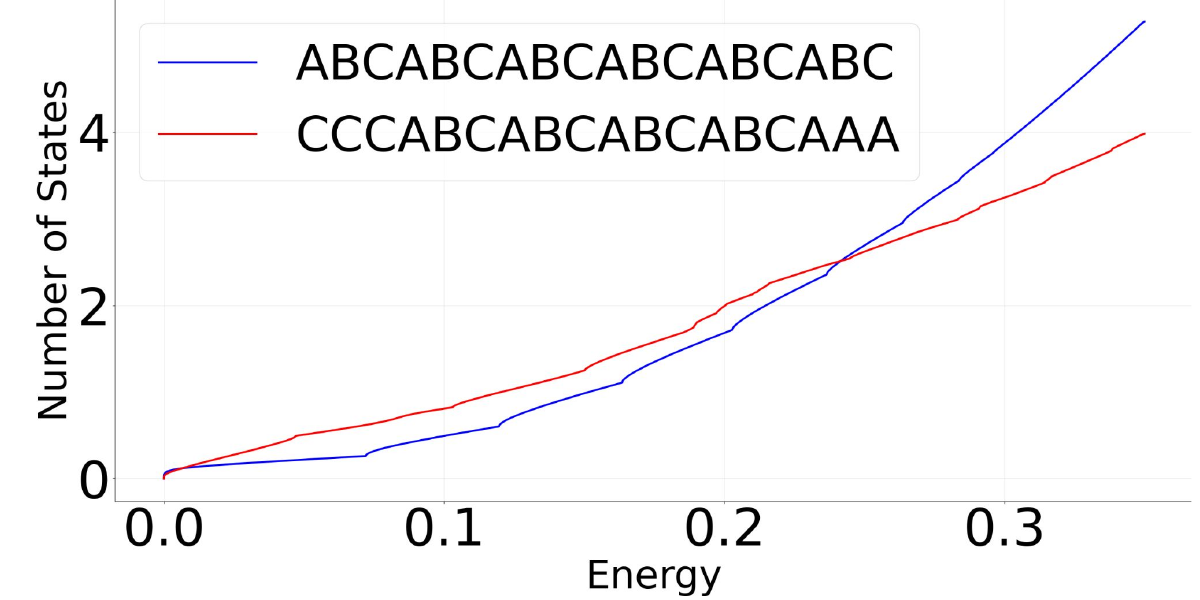}
    \caption{}
    \label{fig:6f}
  \end{subfigure}
  \hfill
  \begin{subfigure}{0.33\textwidth}
    \includegraphics{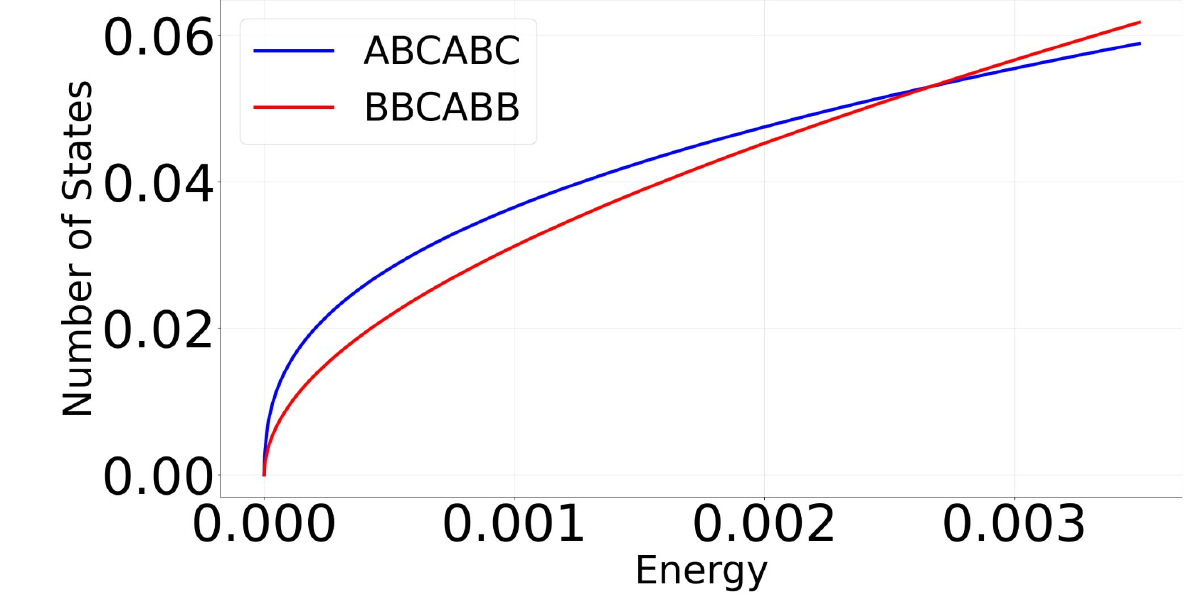}
    \caption{}
    \label{fig:6g}
  \end{subfigure}
  \begin{subfigure}{0.32\textwidth}
    \includegraphics{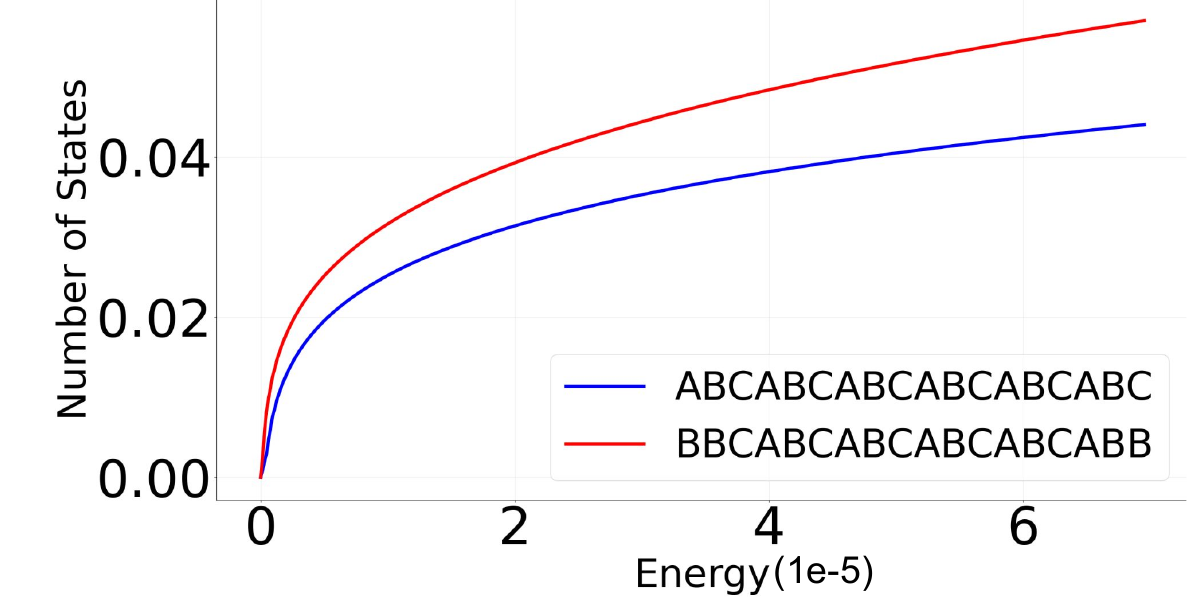}
    \caption{}
    \label{fig:6h}
  \end{subfigure}
  \begin{subfigure}{0.33\textwidth}
    \includegraphics{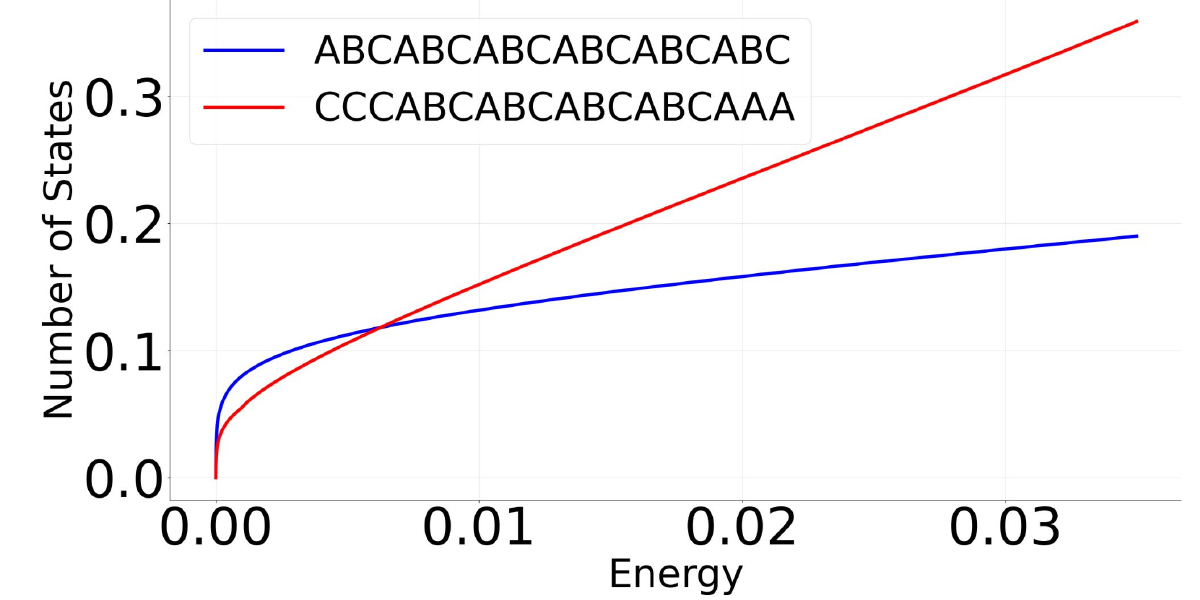}
    \caption{}
    \label{fig:6i}
  \end{subfigure}
  \caption{Comparisons between (a)-(c) DOS calculations with Gaussian smearing $\sigma = \frac{t_\perp}{300}$, (d)-(f) IDOS calculations, and (g)-(i) zoomed in IDOS calculations for rhombohedral stacking with parallel faults (red) and pure rhombohedral stacking (blue) with same stacking length; for (a),(d),(g), $N$ = 6 and we have a BB parallel fault at each end; for (b), (e), (h), $N$ = 18 and we have a BB parallel fault at each end; for (c), (f), (i), $N$ = 18 and we have a CCC parallel fault at each end. We have omitted showing specific stacking orders in the DOS plot due to lack of space, but they are shown in the IDOS plots and the colors are the same for all plots. We see that the density near the Fermi level is enhanced by adding parallel faults to the rhombohedral stacking.}
  \label{fig:6}
\end{figure*}

We see also in Fig. \ref{fig:4a} that embedded ABC substack can induce flatness into these points. We thus have the viability to greatly increase the density of states at the Fermi energy by combining both parallel and rhombohedral stackings. Notably, we will at least be able to double the radius of the region in which our dispersion relation is flat, since the maximum of Eq. \eqref{eq:all_p_value} as our lower bound implies that in the large $N$ limit
\begin{equation}
    \Delta|\vec{p}| \ge \frac{2t_\perp}{v},
\end{equation}
which is two times that of Eq. \eqref{eq:ABC_band_radius} pertaining to pure ABC stacking. To concretely show extended flatness, we plot the Density of States (DOS) and Integrated Density of States (IDOS), which quantifies the number of states in a given energy range, near the Fermi level. Specifically, the DOS is defined as
\begin{equation}
        D(E) = \sum_{i=1}^{2N}\int_{BZ}\delta(\Delta_i(\vec{k}) - E)d\vec{k},
\end{equation}
where $\Delta_i$ is the dispersion relation for the $i$th eigenvalue of our Hamiltonian, and we integrate over the Brillouin zone. The IDOS is defined as
\begin{equation}
    \mu(E) = \int_{0}^ED(\epsilon)d\epsilon
\end{equation}
and we only consider $E>0$ since, in the nearest neighbor model, we have positive and negative energy symmetry. By flatter bands, we expect higher DOS at the energy of the flat band. Moreover, a higher IDOS for low $E$ indicates a greater number of states near the Fermi level. For DOS plots, we use Gaussian smearing with a given $\sigma$ parameter to reduce numerical error, especially since the DOS of a flat band is divergent. For IDOS plots, we do not apply a filter since the integrals converge. Our results are plotted in Fig.~\ref{fig:6}.
Overall, we see a viability to increases state number near the Fermi level. Not only is the zero energy flat band enhanced in Fig.~\ref{fig:6b}, but we also observe the shifting of external flat bands (peaks in the DOS) closer to the Fermi level in Figs.~\ref{fig:6a}-\ref{fig:6c}. The IDOS within $\Delta E \approx 0.3$ of the Fermi level is higher for rhombohedral stacking with $N_i = 2$ parallel embeds, as seen in Figs.~\ref{fig:6d} and ~\ref{fig:6e}. Fig.~\ref{fig:6c} shows how increasing $N_i$ raises this difference, but lower the energy range for which this advantage exists (for faults of length $3$, $\Delta E \approx 0.25$). This shows how parallel embeds shifts the states nearer to the Fermi level. We see that for small stacking length $N$, for example in ~\ref{fig:6a}, ~\ref{fig:6d}, ~\ref{fig:6g}, parallel stacking faults decrease the DOS very close to the Fermi level, as compared with pure rhombohedral stacking of the same length; this is expected since for these low stacking lengths, the change to the main flat band at the Fermi level decreases total flatness despite emergent flat bands away from the Fermi level. Increasing stacking length appears to eliminate this discrepancy. We expect that for higher stacking lengths, an addition of a single parallel fault does not greatly affect the main flat band, and therefore the flat bands outside of the Fermi level with locations independent of total stacking length, as given by Eq. \ref{eq:all_p_value}, contribute to a net increase in DOS at and near the Fermi level.

\begin{figure}
  \setkeys{Gin}{width=\linewidth}
  \begin{subfigure}{.44\textwidth}
    \includegraphics{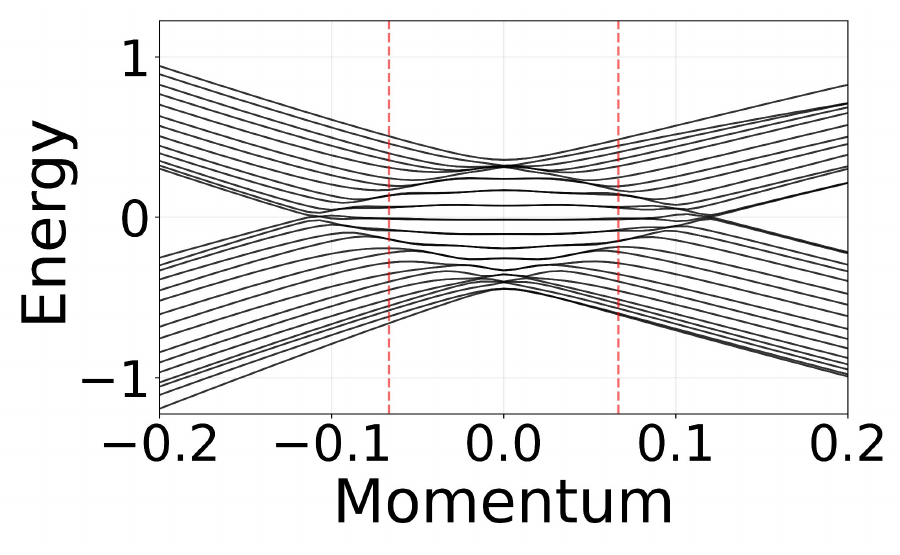}
    \caption{}
    \label{fig:7a}
  \end{subfigure}
  \hfill
  \begin{subfigure}{0.44\textwidth}
    \includegraphics{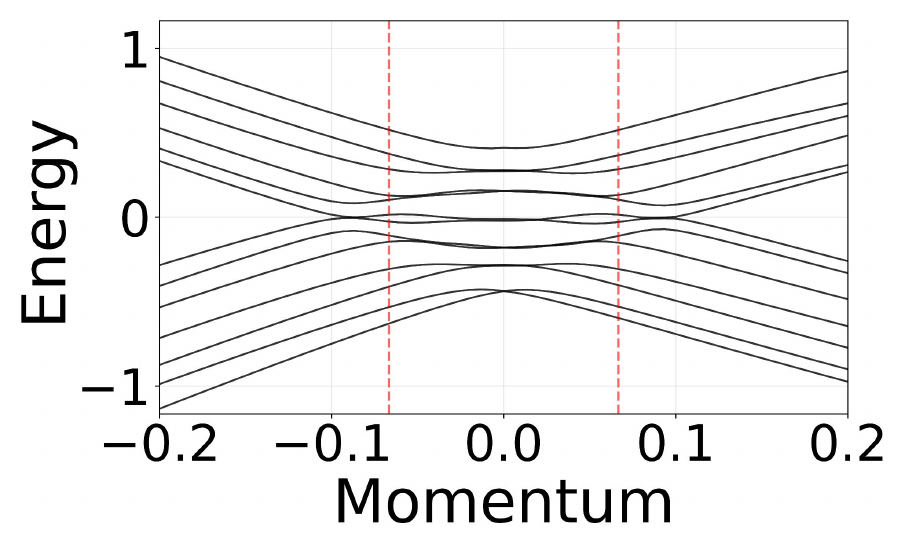}
    \caption{}
    \label{fig:7b}
  \end{subfigure}
  \begin{subfigure}{0.44\textwidth}
    \includegraphics{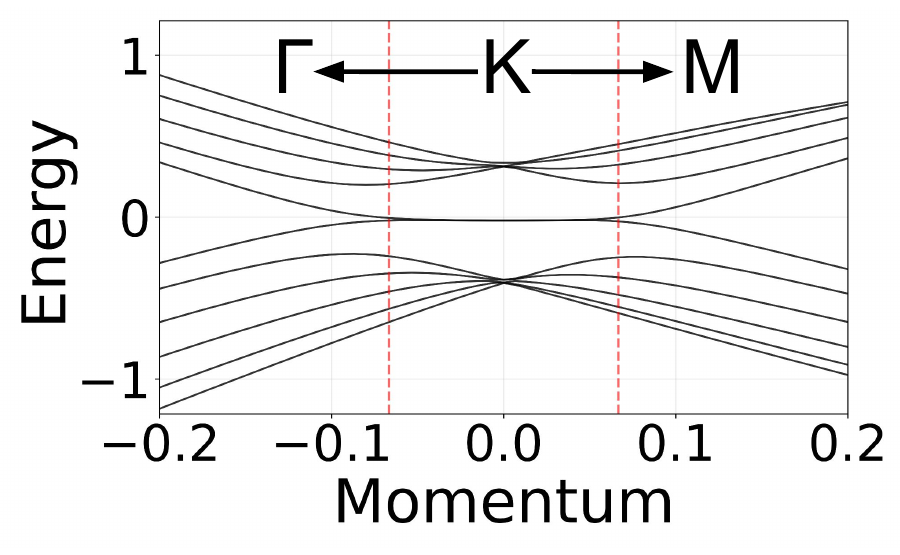}
    \caption{}
    \label{fig:7c}
  \end{subfigure}
  \captionsetup{justification=raggedright, singlelinecheck=false}
  \caption{DFT calculations for (a) AAAAAABCABCCCCCC, (b) AAABCCC, (c) ABCABC stackings. We have plotted our predicted rhombohedral flat band limit as given by Eq. \ref{eq:ABC_band_radius} in red. Notice how (a) and (b) have flat bands well outside this limit. Moreover, we have plotted pure 6-layer rhombohedral stacking in (c), including also the path in the first Brillouin zone, specifically setting the $K$ point as our zero momentum, and plotting along the $\Gamma \to K \to M$ path. Note that we do not hit the $\Gamma$ nor $M$ points in our graphs because we are still focusing on momentum near $K$. }
\end{figure}

Our derivation was done for nearest neighbor hopping only. Our interpretation of rings around the $K$ point is only valid for the region very close to the $K$ or $K'$ point. Away from the $K$ or $K'$ points, the contour plane of the energy surface of graphene exhibits trigonal distortions \cite{Garcia_Ruiz_2021}. It is, however, still maintained that properties, such as extending the flat band, are present in the dispersion relation. Moreover, the similarities between Eq. \ref{eq:all_p_value} and \ref{eq:zero_AA_eigen} suggests the rigidity of substack properties within the whole graphene stack. To show this, we have plotted the various band dispersions calculated from DFT near the $K$ points, along the $\Gamma$ to $K$ to $M$ path as shown in the Figs. \ref{fig:7a}, \ref{fig:7b}, and \ref{fig:7c}.
For all plots, we have included the predicted ABC flat band limit (for large $N$) in red. In \ref{fig:7a} and \ref{fig:7b}, we have plotted AAAAAABCABCCCCCC and AAABCCC, respectively. Clearly, the flat bands of these stacking orders extend far beyond the flat band limit, as we predict by band crossings in the nearest neighbor model. In \ref{fig:7c}, we have plotted ABCABC stacking for reference. We note how the flat band is much more confined to the limits provided by \ref{eq:ABC_band_radius}. We also note that the locations of these crossings roughly align with the nearest neighbor model governed by Eq. \ref{eq:all_p_value}. So, along this path, the nearest neighbor predicted band crossings align well with more accurate calculations. We also mention that other properties of rhombohedral graphene due to long range interactions, including broken symmetry about the Fermi level, are observed in DFT calculations, but with the same asymmetry for the flat bands at $K$ and away from $K$, so that our extended flat bands maintain a similar energy level.

With our arbitrary stacking structures having shown to increase the flatness and number of states near the Fermi level, the ability to engineer a graphene stack with effective and enhanced properties from original graphene stacks with flat electronic bands is likely. It is well understood that the flat band of the rhombohedral graphene is the origin of its many useful properties, including superconductivity and ferromagnetism, in general being indicative of correlated electronic behavior \cite{Henck_2018, guo2025flatbandsurfacestate}. Being able to extend the flat band, and in doing so increase its flatness, suggests that by incorporating stacking faults, we can improve upon these properties and enhance their effectiveness. Indeed, the signature band crossings of parallel stacking serve as an enhancement of the flat band, which could be important to harness graphene's full physical capabilities. It is possible that these enhancements by utilizing stacking faults could, for instance, also be exploited in other graphene based systems, including the twisted bilayer.

\section{Conclusion}\label{sect:conclusion}

We have studied the electronic properties of graphene by considering the properties of the special cases of graphene stacking, AA, AB, and ABC, and applying them to a general, arbitrarily stacked multilayer graphene. We observe a remarkable level of correspondence when we incorporate these stacks together in an arbitrary fashion. Specifically, in studying stacking faults we observed that the embedding of an ABC stack could generate a flat band that very much resembles the sole ABC stack's flat band. The possibility to induce these flat bands which generate correlated electrons while also maintaining linear bands which generate massless Dirac fermions has also been discussed. Moreover, we saw that an embedding of an AA stack allowed us to analytically solve the momenta for which there is an energy eigenvalue of zero. This then allowed us to make a more general statement about the zero energy momenta for arbitrary stacking. The existence of these properties suggests the possibility for a more general consideration of arbitrary stacking. In general, such arbitrary stacking could contain a multitude of parallel, Bernal, and rhombohedral stacking for which one could attempt to predict the properties thereof. We have done so for the momenta for which the energy eigenvalue is zero, which we have found out is in direct relation to the existence of AA substacks; it is therefore plausible that more properties could be associated with the existence of other types of substacks. A primitive method which is what we have shown to be successful in many situations would be to decompose the graphene stack in a well defined manner and considering the decomposition's properties to then predict that of the arbitrary stack; it is qualitatively accurate for the detection of flat bands, and quantitatively accurate for finding the rings of zero energy. These calculations furthermore show the ability to extend the flat band of graphene, which is promising for enhancing its many notable properties.

\nocite{*}

\bibliography{references}

@article{bandtwist,
   title={Band structure of twisted bilayer graphene: Emergent symmetries, commensurate approximants, and Wannier obstructions},
   volume={98},
   ISSN={2469-9969},
   url={http://dx.doi.org/10.1103/PhysRevB.98.085435},
   DOI={10.1103/physrevb.98.085435},
   number={8},
   journal={Physical Review B},
   publisher={American Physical Society (APS)},
   author={Zou, Liujun and Po, Hoi Chun and Vishwanath, Ashvin and Senthil, T.},
   year={2018},
   month=aug }

@article{mactwist,
   title={Graphene bilayers with a twist},
   volume={19},
   ISSN={1476-4660},
   url={http://dx.doi.org/10.1038/s41563-020-00840-0},
   DOI={10.1038/s41563-020-00840-0},
   number={12},
   journal={Nature Materials},
   publisher={Springer Science and Business Media LLC},
   author={Andrei, Eva Y. and MacDonald, Allan H.},
   year={2020},
   month=nov, pages={1265–1275} }

@misc{experimental,
      title={Experimental review of graphene}, 
      author={Daniel R. Cooper and Benjamin D'Anjou and Nageswara Ghattamaneni and Benjamin Harack and Michael Hilke and Alexandre Horth and Norberto Majlis and Mathieu Massicotte and Leron Vandsburger and Eric Whiteway and Victor Yu},
      year={2011},
      eprint={1110.6557},
      archivePrefix={arXiv},
      primaryClass={cond-mat.mes-hall},
      url={https://arxiv.org/abs/1110.6557}}

@book{Kittel,
    author = {Kittel, C.},
    year = {2004},
    title = {Introduction to Solid State Physics},
    publisher = {John Wiley and Son},
    edition = {8}}

@article{SSH,
   title={Analytical solution of open crystalline linear 1D tight-binding models},
   volume={53},
   ISSN={1751-8121},
   url={http://dx.doi.org/10.1088/1751-8121/ab6a6e},
   DOI={10.1088/1751-8121/ab6a6e},
   number={7},
   journal={Journal of Physics A: Mathematical and Theoretical},
   publisher={IOP Publishing},
   author={Marques, A M and Dias, R G},
   year={2020},
   month=jan, pages={075303}}

@misc{rutgers,
    author = {Roy, M.},
    title = {The Tight Binding Model},
    month = {May},
    year = {2015},
    publisher = {Rutgers University},}

@misc{crystal_structure,
    author = {Gray, D. and McCaughan, A. and Mookerji, B.},
    title = {Crystal Structure of Graphite, Graphene and Silicon},
    month = {March},
    year = {2009},
    publisher = {West Virginia University}
}

@misc{tutorial,
    author = {Oreg, Y.},
    title = {Tutorial 1: Introduction to Graphene},
    month = {April},
    year = {2017},
    publisher = {Weizmann Institute of Science}}

@article{Semenoff,
  title = {Condensed-Matter Simulation of a Three-Dimensional Anomaly},
  author = {Semenoff, Gordon W.},
  journal = {Phys. Rev. Lett.},
  volume = {53},
  issue = {26},
  pages = {2449--2452},
  numpages = {0},
  year = {1984},
  month = {Dec},
  publisher = {American Physical Society},
  doi = {10.1103/PhysRevLett.53.2449},
  url = {https://link.aps.org/doi/10.1103/PhysRevLett.53.2449}}

@article{Macdonald,
   title={Electronic Structure of Multilayer Graphene},
   volume={176},
   ISSN={0375-9687},
   url={http://dx.doi.org/10.1143/PTPS.176.227},
   DOI={10.1143/ptps.176.227},
   journal={Progress of Theoretical Physics Supplement},
   publisher={Oxford University Press (OUP)},
   author={Min, Hongki and MacDonald, Allan H.},
   year={2008},
   pages={227–252} }

@article{Yan,
   title={Enhanced optical conductivity induced by surface states in ABC-stacked few-layer graphene},
   volume={83},
   ISSN={1550-235X},
   url={http://dx.doi.org/10.1103/PhysRevB.83.245418},
   DOI={10.1103/physrevb.83.245418},
   number={24},
   journal={Physical Review B},
   publisher={American Physical Society (APS)},
   author={Yan, Jia-An and Ruan, W. Y. and Chou, M. Y.},
   year={2011},
   month=jun }

@article{Aufbau,
    title = {On the application of quantum theory to atomic structure},
    author = {Bohr, Niels},
    year = {1923},
    journal = {Zeitschrift für Physik},
    volume = {13},
    issue = {1},
    ISSN = {0044-3328},
    url = {https://doi.org/10.1007/BF01328209},
    publisher = {Springer Berlin Heidelberg},
    pages = {117-165},
    doi = {10.1007/BF01328209}}

@article{Schrieffer-Wolff,
  title = {Motion of Electrons and Holes in Perturbed Periodic Fields},
  author = {Luttinger, J. M. and Kohn, W.},
  journal = {Phys. Rev.},
  volume = {97},
  issue = {4},
  pages = {869--883},
  numpages = {0},
  year = {1955},
  month = {Feb},
  publisher = {American Physical Society},
  doi = {10.1103/PhysRev.97.869},
  url = {https://link.aps.org/doi/10.1103/PhysRev.97.869}}

@article{Wallace,
  title = {The Band Theory of Graphite},
  author = {Wallace, P. R.},
  journal = {Phys. Rev.},
  volume = {71},
  issue = {9},
  pages = {622--634},
  numpages = {0},
  year = {1947},
  month = {May},
  publisher = {American Physical Society},
  doi = {10.1103/PhysRev.71.622},
  url = {https://link.aps.org/doi/10.1103/PhysRev.71.622}}

@article{superconductor,
   title={Superconductivity and magnetism in the surface states of ABC-stacked multilayer graphene},
   volume={108},
   ISSN={2469-9969},
   url={http://dx.doi.org/10.1103/PhysRevB.108.144504},
   DOI={10.1103/physrevb.108.144504},
   number={14},
   journal={Physical Review B},
   publisher={American Physical Society (APS)},
   author={Awoga, Oladunjoye A. and Löthman, Tomas and Black-Schaffer, Annica M.},
   year={2023},
   month=oct }

@article{Koshino_nature,
    title = {Electronic properties of stacking faults in Bernal graphite},
    author = {Sarsfield, P.J. and Slizovskiy, S. and Koshino, M. and Fal'ko V.},
    year = {2025},
    journal = {npj Computational Materials},
    volume = {11},
    issue = {1},
    ISSN = {2057-3960},
    url = {https://doi.org/10.1038/s41524-025-01641-2},
    doi = {10.1038/s41524-025-01641-2}}

@article{Castro_Neto,
   title={The electronic properties of graphene},
   volume={81},
   ISSN={1539-0756},
   url={http://dx.doi.org/10.1103/RevModPhys.81.109},
   DOI={10.1103/revmodphys.81.109},
   number={1},
   journal={Reviews of Modern Physics},
   publisher={American Physical Society (APS)},
   author={Castro Neto, A. H. and Guinea, F. and Peres, N. M. R. and Novoselov, K. S. and Geim, A. K.},
   year={2009},
   month=jan, pages={109–162} }

@book{invertable_det, place={Cambridge}, series={Econometric Exercises}, title={Matrix Algebra}, publisher={Cambridge University Press}, author={Abadir, Karim M. and Magnus, Jan R.}, year={2005}, collection={Econometric Exercises}}

@book{Chebyshev,
  author    = {Abramowitz, Milton and Stegun, Irene A.},
  title     = {Handbook of Mathematical Functions with Formulas, Graphs, and Mathematical Tables},
  publisher = {Dover Publications},
  year      = {1964},
  address   = {New York},
}

@article{Chen2019TLG,
  title={Signatures of Gate-Tunable Superconductivity in Trilayer Graphene/Boron Nitride Moir\'e Superlattice},
  author={Chen, Guorui and Sharpe, Aaron L. and Gallagher, Patrick and Rosen, Ilan T. and Fox, Eli and Jiang, Lili and Lyu, Bosai and Li, Hongyuan and Watanabe, Kenji and Taniguchi, Takashi and Jung, Jeil and Shi, Zhiwen and Goldhaber-Gordon, David and Zhang, Yuanbo and Wang, Feng},
  journal={arXiv preprint arXiv:1901.04621},
  year={2019},
  eprint={1901.04621},
  archivePrefix={arXiv},
  primaryClass={cond-mat.mes-hall},
  url={https://arxiv.org/abs/1901.04621}
}

@article{Zhou2021RTG,
  title={Superconductivity in rhombohedral trilayer graphene},
  author={Zhou, Haoxin and Xie, Tian and Taniguchi, Takashi and Watanabe, Kenji and Young, Andrea F.},
  journal={Nature},
  volume={598},
  number={7881},
  pages={434--438},
  year={2021},
  publisher={Nature Publishing Group},
  doi={10.1038/s41586-021-03926-0},
  eprint={2106.07640},
  archivePrefix={arXiv},
  primaryClass={cond-mat.mes-hall},
  url={https://arxiv.org/abs/2106.07640}
}

@article{Park2021MAG,
  title={Magic-Angle Multilayer Graphene: A Robust Family of Moir\'e Superconductors},
  author={Park, Jeong Min and Cao, Yuan and Xia, Liqiao and Sun, Shuwen and Watanabe, Kenji and Taniguchi, Takashi and Jarillo-Herrero, Pablo},
  journal={arXiv preprint arXiv:2112.10760},
  year={2021},
  eprint={2112.10760},
  archivePrefix={arXiv},
  primaryClass={cond-mat.supr-con},
  url={https://arxiv.org/abs/2112.10760}
}

@article{Han2024Chiral,
  title={Signatures of Chiral Superconductivity in Rhombohedral Graphene},
  author={Han, Tonghang and Lu, Zhengguang and Hadjri, Zach and Shi, Lihan and Wu, Zhenghan and Xu, Wei and Yao, Yuxuan and Cotten, Armel A. and Sedeh, Omid Sharifi and Weldeyesus, Henok and Yang, Jixiang and Seo, Junseok and Ye, Shenyong and Zhou, Muyang and Liu, Haoyang and Shi, Gang and Hua, Zhenqi and Watanabe, Kenji and Taniguchi, Takashi and Xiong, Peng and Zumb\"uhl, Dominik M. and Fu, Liang and Ju, Long},
  journal={arXiv preprint arXiv:2408.15233},
  year={2024},
  eprint={2408.15233},
  archivePrefix={arXiv},
  primaryClass={cond-mat.mes-hall},
  url={https://arxiv.org/abs/2408.15233}
}

@article{Latil2006FLG,
  title={Charge Carriers in Few-Layer Graphene Films},
  author={Latil, Sylvain and Henrard, Luc},
  journal={Physical Review Letters},
  volume={97},
  number={3},
  pages={036803},
  year={2006},
  publisher={American Physical Society},
  doi={10.1103/PhysRevLett.97.036803}
}

@article{Koshino_2013,
   title={Multilayer graphenes with mixed stacking structure: Interplay of Bernal and rhombohedral stacking},
   volume={87},
   ISSN={1550-235X},
   url={http://dx.doi.org/10.1103/PhysRevB.87.045420},
   DOI={10.1103/physrevb.87.045420},
   number={4},
   journal={Physical Review B},
   publisher={American Physical Society (APS)},
   author={Koshino, Mikito and McCann, Edward},
   year={2013},
   month=jan }

@article{Garcia_Ruiz_2021,
   title={Full Slonczewski-Weiss-McClure parametrization of few-layer twistronic graphene},
   volume={104},
   ISSN={2469-9969},
   url={http://dx.doi.org/10.1103/PhysRevB.104.085402},
   DOI={10.1103/physrevb.104.085402},
   number={8},
   journal={Physical Review B},
   publisher={American Physical Society (APS)},
   author={Garcia-Ruiz, Aitor and Deng, Hai-Yao and Enaldiev, Vladimir V. and Fal’ko, Vladimir I.},
   year={2021},
   month=aug }

@article{Henck_2018,
   title={Flat electronic bands in long sequences of rhombohedral-stacked graphene},
   volume={97},
   ISSN={2469-9969},
   url={http://dx.doi.org/10.1103/PhysRevB.97.245421},
   DOI={10.1103/physrevb.97.245421},
   number={24},
   journal={Physical Review B},
   publisher={American Physical Society (APS)},
   author={Henck, Hugo and Avila, Jose and Ben Aziza, Zeineb and Pierucci, Debora and Baima, Jacopo and Pamuk, Betül and Chaste, Julien and Utt, Daniel and Bartos, Miroslav and Nogajewski, Karol and Piot, Benjamin A. and Orlita, Milan and Potemski, Marek and Calandra, Matteo and Asensio, Maria C. and Mauri, Francesco and Faugeras, Clément and Ouerghi, Abdelkarim},
   year={2018},
   month=jun }

@misc{guo2025flatbandsurfacestate,
      title={Flat band surface state superconductivity in thick rhombohedral graphene}, 
      author={Yi Guo and Owen I. Sheekey and Trevor Arp and Kryštof Kolář and Thibault Charpentier and Ludwig Holleis and Ben Foutty and Aidan Keough and Maya Kang-Chou and Martin E. Huber and Takashi Taniguchi and Kenji Watanabe and Cyprian Lewandowski and Andrea F. Young},
      year={2025},
      eprint={2511.17423},
      archivePrefix={arXiv},
      primaryClass={cond-mat.supr-con},
      url={https://arxiv.org/abs/2511.17423}, 
}

@article{pwscf,
   title={QUANTUM ESPRESSO: a modular and open-source software project for quantum simulations of materials},
   volume={21},
   ISSN={1361-648X},
   url={http://dx.doi.org/10.1088/0953-8984/21/39/395502},
   DOI={10.1088/0953-8984/21/39/395502},
   number={39},
   journal={Journal of Physics: Condensed Matter},
   publisher={IOP Publishing},
   author={Giannozzi, Paolo and Baroni, Stefano and Bonini, Nicola and Calandra, Matteo and Car, Roberto and Cavazzoni, Carlo and Ceresoli, Davide and Chiarotti, Guido L and Cococcioni, Matteo and Dabo, Ismaila and Dal Corso, Andrea and de Gironcoli, Stefano and Fabris, Stefano and Fratesi, Guido and Gebauer, Ralph and Gerstmann, Uwe and Gougoussis, Christos and Kokalj, Anton and Lazzeri, Michele and Martin-Samos, Layla and Marzari, Nicola and Mauri, Francesco and Mazzarello, Riccardo and Paolini, Stefano and Pasquarello, Alfredo and Paulatto, Lorenzo and Sbraccia, Carlo and Scandolo, Sandro and Sclauzero, Gabriele and Seitsonen, Ari P and Smogunov, Alexander and Umari, Paolo and Wentzcovitch, Renata M},
   year={2009},
   month=sep, pages={395502} }
\end{document}